\documentclass[journal,twoside]{IEEEtran}

\usepackage{amsmath}
\usepackage{graphicx}
\usepackage{algorithmic}
\usepackage{algorithm}
\usepackage{hyperref}
\usepackage{verbatim}
\usepackage{enumerate}
\usepackage{amsfonts}
\usepackage{amsthm, amssymb}
\usepackage{epstopdf}
\usepackage{subfig}
\usepackage{lineno}
\usepackage{tikz}
\usetikzlibrary{calc,trees,positioning,arrows,chains,shapes.geometric,
    decorations.pathreplacing,decorations.pathmorphing,shapes,
    matrix,shapes.symbols}
\tikzset{
>=stealth',
  punktchain/.style={
    rectangle, 
    rounded corners, 
     fill=black!10,
    draw=black, very thick,
    text width=20em, 
    minimum height=1em, 
    text centered, 
    on chain},
  line/.style={draw, thick, <-},
  element/.style={
    tape,
    top color=white,
    bottom color=blue!50!black!60!,
    minimum width=8em,
    draw=blue!40!black!90, very thick,
    text width=10em, 
    minimum height=3.5em, 
    text centered, 
    on chain},
  every join/.style={->, thick,shorten >=1pt},
  decoration={brace},
  tuborg/.style={decorate},
  tubnode/.style={midway, right=2pt},
}

\begin{document} 
\title{Component Based Modeling of Ultrasound Signals}
\author{Yael Yankelevsky, Zvi Friedman, and Arie Feuer,~\IEEEmembership{Fellow,~IEEE}}
\maketitle

\begin{abstract}
This work proposes a component based model for the raw ultrasound signals acquired by the transducer elements.
Based on this approach, before undergoing the standard digital processing chain, every sampled raw signal is first decomposed into a smooth background signal and a strong reflectors component.
The decomposition allows for a suited processing scheme to be adjusted for each component individually. 
We demonstrate the potential benefit of this approach in image enhancement by suppressing side lobe artifacts, and in improvement of digital data compression.
Applying our proposed processing schemes to real cardiac ultrasound data, we show that by separating the two components and compressing them individually, over twenty-fold reduction of the data size is achieved while retaining the image contents.
\end{abstract}


\section{Introduction}
Medical ultrasound imaging allows visualization of internal body structures by radiating them with acoustic energy and analyzing the returned echoes. The two-dimensional image typically comprises of multiple one-dimensional scan lines, each constructed by integrating the data collected by the transducer elements following the transmission of an acoustic pulse along a narrow beam. As the transmitted pulse propagates through the body, echoes are scattered by acoustic impedance perturbations in the tissue. These back-scattered echoes are detected by the transducer elements and combined, after aligning them with the appropriate time delays, in a process referred to as beamforming, which results in Signal-to-Noise Ratio (SNR) enhancement. Each resulting beamformed signal forms a line in the image.

The processing of ultrasound signals or images is often based on an underlying signal model.
Resulting from interferences of randomly spaced scatterers, the ultrasonic echoes have a stochastic nature which is manifested in the resulting image as speckle noise.
In the context of denoising or despeckling, various statistical models for speckle have been proposed that are inspired by the physical process underlying speckle formation.
A commonly used model for the ultrasonic echo assumes that any range cell contains a large number of randomly located scatterers, hence the envelope of the echo obeys a Rayleigh distribution \cite{Burckhardt1978}.
When the number of scatterers is not large enough or the scatterers are not randomly located, deviations from the Rayleigh model may occur. To account for such deviations, other models have been proposed, such as the Rician distribution \cite{Wagner1983, Insana1985}, K-distribution \cite{Jakeman1987}, and others. 

Beyond the statistical models, several works (e.g. \cite{Girault1998}, \cite{Michailovich2005}) are based on parametric spectral analysis and model the ultrasonic radio frequency (RF) signal as an autoregressive process (AR), i.e. as an output of a linear filter driven by a white Gaussian noise.

Another suggested model \cite{Taxt1995,Michailovich2006} considers the ultrasound RF image to be a result of the convolution of the tissue reflectivity function with the two-dimensional point-spread function (PSF) of the imaging system.

Some other less common models have also been proposed, such as a local polynomial representation of the ultrasonic pulse log-spectrum \cite{Adam2002} or modeling of the RF echo as a power-law shot-noise process \cite{Kutay2001}.

A more popular approach is based on a sparsity assumption, i.e. considering the signals to be sparse under some basis or dictionary. 
Explicitly, one seeks an approximation $\hat{\mathbf{y}}=\mathbf{D\hat{x}}$ of the signal $\mathbf{y}\in\mathbb{R}^N$ that obeys 
\begin{equation}
\mathbf{\hat{x}}=\arg\underset{\mathbf{x}}{\min}\;\|\mathbf{y}-\mathbf{Dx}\|_2^2\quad \mbox{ subject to }\quad \|\mathbf{x}\|_0\leq T_0
\end{equation}
where $\mathbf{x}\in\mathbb{R}^K$ is the sparse representation vector, $\|\mathbf{x}\|_0$ is the $\ell_0$ pseudo-norm counting the non-zero entries in the vector $\mathbf{x}$, and the matrix $\mathbf{D}\in\mathbb{R}^{N\times K}$ is the representation dictionary. This dictionary may be either predetermined or learned from the data itself, in a process referred to as dictionary learning. The underlying idea is an adaptive adjustment of the dictionary to observed signal characterizations, by solving
\begin{equation}
\arg\underset{\mathbf{D},\mathbf{X}}{\min}\|\mathbf{Y}-\mathbf{DX}\|_F^2\quad\mbox{ subject to }\quad \|\mathbf{x_i}\|_0\leq T_0 \quad \forall i
\end{equation}
where $\mathbf{Y}\in\mathbb{R}^{N\times M}$ is a data matrix having the observed signals at its columns, and $\mathbf{x_i}$ is the $i$-th column of the coefficients matrix $\mathbf{X}\in\mathbb{R}^{K\times M}$.

Based on the sparsity model, Quinsac et al. \cite{Quinsac2010, Quinsac2012} suggested using Fourier or wavelets as the representation basis.
Friboulet et al. \cite{Friboulet2010, Liebgott2013} used a dictionary of directional wave atoms, which shows good properties for sparsely representing warped oscillatory patterns. The authors showed that following random sparse sampling, reconstruction using the wave atoms dictionary yields better results compared with the Fourier or Wavelets bases.

Another set of works~ \cite{Tur2010,Wagner2012} assumed sparsity by modeling the ultrasonic echo as a finite stream of strong pulses. This model suggested that the echo consists of $L$ pulses that are replicas of a known-shape pulse with unknown time-shifts and amplitudes:
\begin{equation} \label{Eq:pulseModel}
y(t)=\sum_{l=1}^L a_lh(t-t_l)
\end{equation}

Few works attempted to select the representation basis for the ultrasound images (or image patches) using dictionary learning methods \cite{Tosic2010, Lorintiu2013, Lorintiu2014}.
To our knowledge, no similar attempts were made to apply dictionary learning approaches to one-dimensional ultrasound signals.
\newline

In this work, we propose a new model for raw ultrasonic signals as the sum of two components, both carrying valuable information: the strong reflectors, that are highly important for tracking purposes in cardiac ultrasound imaging \cite{Trahey1987, Trahey1991}, and the speckle, also referred to in this work as the background component, that characterizes the microscopic structure of the tissue \cite{Burckhardt1978}.

For the strong reflectors we assume a stream of pulses model, similar to the one proposed by \cite{Tur2010, Wagner2012}.
For the background component, we learn a suitable dictionary using the K-SVD method \cite{Aharon2006}.
As we show next, the decomposition at an early stage allows for a suited processing scheme to be adjusted for each component individually.
We demonstrate the potential benefits of this approach in addressing two of the issues that stand as disadvantages of ultrasound imaging, particularly the large amount of data needed to be stored and processed and the inherent side lobes artifacts.
\newline

Throughout this paper, we use the term \emph{raw signal} to refer to a signal acquired and sampled by a single transducer element for a single given scan line, i.e. before combining them in the digital beamforming process.
\newline

The rest of the paper is organized as follows: 
In Section~\ref{section:Decomposition} we describe two proposed methods for decomposing the raw signals into the strong reflectors and background components.

We then demonstrate the advantage of this model for data compression. 
In Section~\ref{section:BackgroundCompression} and Section~\ref{section:compressedBFmain} we describe the modified processing chain of the background components, consisting of their sparse representation over a trained dictionary followed by their integration in a modified beamforming process that is applied directly at the representation domain.
In Section~\ref{section:SideLobeSuppression} we present a slightly modified decomposition approach that achieves better suppression of the side lobes artifacts.
Section~\ref{section:SimulationResults} then presents simulation and experimental results.
We summarize the paper in Section~\ref{section:Conclusions} with some concluding remarks.

\section{Signal Decomposition} \label{section:Decomposition}

As stated in the introductory section, our model assumes each raw signal to be composed of a background signal and a strong reflectors component. We shall therefore begin by decomposing each such signal to its individual components, denoted $y_b$, $y_s$ respectively. 
The decomposition algorithm is based on a greedy detection of the strong reflectors followed by their separation from the original signal.
We here propose two methods for performing the decomposition: one in the time-frequency domain and the other using the in-phase and quadrature representation.

\subsection{Strong Reflectors Removal via Short-Time Fourier Transform} \label{section:STFTdecomposition}

Modeling the strong reflectors component, we mostly adopt the "stream of pulses" model given by Equation~\eqref{Eq:pulseModel}, according to which this component is composed of a limited number of strong pulses, that are amplified and delayed replicas of a known-shape pulse.
This pulse has the form of a sinusoid signal oscillating at the transmission frequency $f_0$ in a Gaussian envelope.

As an extension to the previously proposed model, we suggest that the returning pulse shape is somewhat corrupted with respect to the transmitted pulse. This corruption may be manifested in either a frequency shift, resulting from the frequency dependent attenuation \cite{Azhari2010}, or a phase shift formed between the carrier wave and the Gaussian envelope.

In order to account for those possible corruptions, we propose to represent the strong reflectors in a time-frequency domain, using the Short-Time Fourier Transform (STFT).
This will allow simultaneous optimization of both the time-delay and frequency shift.

Denoting the STFT of $y(t)$ by $\mathbf{Y}(t,\omega)$, the STFT decomposition is
\begin{equation}
\begin{aligned}
\mathbf{Y}(t,\omega) &= \mathbf{Y_b}(t,\omega)+\mathbf{Y_s}(t,\omega) \\ 
&=  \mathbf{Y_b}(t,\omega)+\sum_{k=1}^{L}{a_k\mathbf{H}(t-t_k,\omega-\omega_k)}
\end{aligned}
\end{equation}

The proposed algorithm for STFT-based separation of the strong reflectors is described in Algorithm~\ref{Alg:STFTdecomposition}.
Upon reaching the stopping condition, we are left with two separate signals in the time-frequency domain. By applying inverse STFT, the time-domain components can be reconstructed.

\begin{algorithm}[ht]
\caption{STFT-based Decomposition}
\label{Alg:STFTdecomposition}
\begin{algorithmic}[]
\STATE \textbf{Task:} Decompose a given signal $y$ into the strong reflectors and background components ($y_s,y_b$ respectively)
\STATE \textbf{Inputs:} The signal's STFT $\mathbf{Y}(t,\omega)$, the STFT signature of the pulse model $\mathbf{H}(t,\omega)$ centered such that $\arg\underset{(t,\omega)}{\max}|\mathbf{H}(t,\omega)|=(0,0)$, the maximal number of pulses $L$, and an error threshold $\epsilon_0$.
\STATE \textbf{Initialization:} Set the initial residual $\mathbf{r^0} = \mathbf{Y}(t,\omega)$
\STATE \textbf{Main Iteration:}
for $k=1,...,L$ perform the following:
\begin{itemize}
	\item Locate the strongest reflection $(t_k,\omega_k)$ with magnitude $a_k$:
	\begin{equation} \nonumber
	(t_k,\omega_k)=\arg\underset{(t,\omega)}{\max}|\mathbf{Y}(t,\omega)| \quad ; \quad a_k = \mathbf{Y}(t_k,\omega_k)
	\end{equation}
	\item Residual update: $\mathbf{r^k} = \mathbf{r^{k-1}}-a_k\mathbf{H}(t-t_k,\omega-\omega_k)$
	\item Stopping Rule: If $\|\mathbf{r^k}\|_\infty<\epsilon_0$, stop. Otherwise, apply another iteration.
\end{itemize}
\STATE \textbf{Output:} 
$y_s(t) = ISTFT\left(\sum_{j=1}^{k}{a_j\mathbf{H}(t-t_j,\omega-\omega_j)}\right)$, $y_b(t) = ISTFT(\mathbf{r^k})$.
\end{algorithmic}
\end{algorithm}

In the presented algorithm, the strong reflectors are detected as the maximal magnitude peaks of the STFT. In practice, the detection can be improved by matching the known pulse pattern in a narrow region around each peak.
The maximal number of pulses $L$ and error threshold $\epsilon_0$ are chosen empirically.

It can be observed that the strong reflectors are naturally compressed by saving only the pulse model parameters $\{a_k,t_k,\omega_k\}_{k=1}^L$ along the decomposition process.
This may be thought of as sparse coding over a very large dictionary whose atoms represent all the possible time and frequency shifts of the known pulse.
However, due to the high sampling rate of the signals and the enormous dimensions of such dictionary, standard pursuit techniques like OMP are not feasible and an alternative amplitude-based pulse matching was here performed.

\subsection{Strong Reflectors Removal in I/Q} \label{section:IQdecomposition}
The ultrasonic RF signal may be modeled as a complex, low-frequency baseband signal, modulated by a carrier wave that oscillates at the much higher transmission frequency $\omega_0=2\pi f_0$. 
The resulting RF signal takes the form 
\begin{equation} \label{Eq:RF}
	y_{RF}(t)= I(t)\cos(\omega_0 t)-Q(t)\sin(\omega_0 t)
\end{equation} 
and the baseband signal is assumed to be of the form
\begin{equation} \label{Eq:IQ}
	y_{IQ}(t)=I(t)+jQ(t)
\end{equation}
where $I(t),Q(t)$ are slowly varying (in comparison with $f_0$) In-phase and Quadrature components, and $t$ is the time along the beam. 
IQ demodulation is therefore the process of extracting the IQ signal \eqref{Eq:IQ} from the measured physical RF signal \eqref{Eq:RF}.

Since the baseband components I/Q change slowly, they can be sampled at a lower frequency, thus reducing the amount of data to be saved.
Utilizing this fact, modern ultrasound systems sample the complex baseband signal after IQ demodulation instead of directly sampling the received RF signal. 
Besides the advantage of a lower sampling rate, identifying the strong reflections at signals of lower frequency is less bound to errors such as time inaccuracies or phase shifts. 
Consequently, it would be beneficial to accommodate the proposed decomposition to such systems and detect the strong reflectors at the I/Q components.

Working in baseband, there is no longer a need to match the full pulse model (including the frequency and phase estimation) and only a Gaussian envelope should be adjusted with the appropriate delay and amplitude.

The algorithm performed for each raw signal independently is generally similar to the one implemented in STFT, with few differences:
The inputs are the baseband signal $\mathbf{\varphi_{IQ}(t)}$ and the modeled pulse envelope $\mathbf{g(t)}$, instead of the STFT of the signal and pulse. Respectively, the residual is initialized with $\mathbf{r^0} = \mathbf{\varphi_{IQ}(t)}$ and updated by subtracting the adjusted envelope from I and Q individually: 
\begin{equation}\nonumber
\bold{r_{I,Q}^k} = \bold{r_{I,Q}^{k-1}}-a_{I,Q}^k\bold{g(t-t_k)}
\end{equation} 
Additionally, the saved parameters for the strong reflectors now contain only the time delays $\left\lbrace t_j\right\rbrace_{j=1}^{k}$ and respective amplitudes, and the final residual $\mathbf{r^k}$ represents the baseband background component.

\section{Data Compression} \label{section:BackgroundCompression}
According to the Shannon-Nyquist sampling theorem, the sampling rate at each transducer element should be at least twice the bandwidth of the detected signal. Taking into account the high frequency used for ultrasound imaging, the number of transducer elements and the number of lines in an image, the amount of data needed to be transferred and processed is very large, motivating methods to reduce the amount of needed data without compromising the reconstructed image quality and its diagnostic credibility.

In state-of-the-art ultrasound systems, an initial reduction of data size is already made possible by exploiting the fact that the detected signals are modulated onto a carrier and occupy only a portion of the entire baseband bandwidth. Therefore by demodulating the received signals, the sampling rate, and consequently the data size, may be reduced. Nevertheless, this reduction is far from fulfilling the compression potential. 

To further reduce the amount of data, several approaches were proposed in the literature.
Several works have recently adapted compressed sensing (CS) methods to ultrasound signals (e.g. \cite{Quinsac2010,Quinsac2012,Basarab2013,Friboulet2010,Liebgott2012,Liebgott2013,Schiffner2011,Lorintiu2013}). 
Other works (e.g. \cite{Li2012}) treat the collection of raw signals detected by the transducer elements for a single scan line as a two-dimensional image, and attempt to compress it using standard techniques such as JPEG compression.

Another work attempting to reduce the amount of sampled data in ultrasound signals, based on complementary ideas arising from the Finite Rate of Innovation (FRI) framework \cite{Vetterli2002,Blu2008}, was carried out by Tur et al. \cite{Tur2010}. The authors consider the returning echo to be a stream of pulses that are replicas of a known-shape pulse with unknown time-shifts and amplitudes.
Assuming that overall $L$ pulses were reflected back to the transducer from the pulse's propagation path, the detected signal is completely defined by $2L$ degrees of freedom, corresponding to the unknown time delays and amplitudes of these pulses.
Based on the FRI framework, these $2L$ parameters are estimated and the signal recovered from a minimal subset of $2L$ of the signal's Fourier series coefficients. 
The needed coefficients are recovered from low rate samples of the analog signals, as the sampling frequency is now determined by the number of pulses $L$, which is rather small compared with the bandwidth of the transmitted pulse, leading to a substantial sample rate reduction.

This result was extended by Wagner et al. \cite{Wagner2012}, that applied concepts of the CS framework in order to directly reconstruct the beamformed signals from low rate samples of the individual signals detected by the different transducer elements.

These works achieve an almost eight-fold reduction of sample rate, however the reconstructed data is partial as it only contains the macroscopic reflections and does not include the speckle.
\newline

A different approach used in radar is based on the CLEAN algorithm \cite{Chira2012}. This algorithm performs an iterative search of the strong reflections, where at each iteration the maximal amplitude of the returned signal is located and the signal is then attenuated around this detected maxima using some attenuation function. This process is repeated until the residual signal reaches a predefined threshold, after which the strong reflectors could be reconstructed from their locations and amplitudes collected across iterations. Like the previous methods, the main shortcoming of this approach is that it disregards speckle and only reconstructs the strong targets.
\newline

In the following, we demonstrate how the proposed decomposition approach may be utilized for improving digital compression of ultrasound signals by altering the processing scheme.

To motivate the decomposition based compression, we first note that the strong reflectors are readily compressible since their number is limited and they can be fully characterized by a few pulse model parameters.  
Resulting from intereferences of weak ultrasonic reflections, speckle is typically characterized by a statistical model with few parameters, indicating that it could also be easily sparsified. 
Therefore, while a joint compressible model may be hard to obtain, a decomposition based approach results in two components that are each highly compressible.

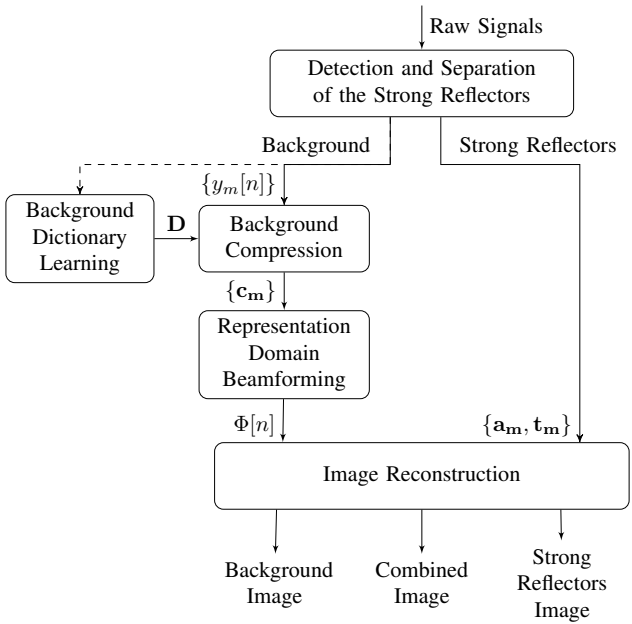
\begin{figure}[h] 
\tikzstyle{block} = [rectangle, draw, fill=blue!0, 
    text width=7em, text centered, rounded corners, minimum height=3em]
\tikzstyle{cloud} = [minimum height=0.6em, text width=4.5em, text centered]
\tikzstyle{line} = [draw, -latex']
\hspace{+0.1cm}
\scalebox{0.84}{
\begin{tikzpicture}[node distance = 1.5cm, auto]

\node [block,text width=13em] (Decomposition) {Detection and Separation of the Strong Reflectors};
\node [above=0.7cm of Decomposition] (raw) {};
\node [below=1.8cm of Decomposition] (dummy) {};
\node [right=1.5cm of dummy] (strong) {};
\node [block, left=0.7cm of dummy,text width=7em] (Compression) {Background Compression};
    \node [block, left=0.7cm of Compression, text width=6em] (KSVD) {Background \\ Dictionary \\ Learning};
    \node [block, below=0.6cm of Compression, text width=7em] (Beamforming) {Representation \\ Domain \\ Beamforming};
    \node [block, below=3.1cm of dummy, text width=18em] (Reconstruction) {Image Reconstruction};
\path [line] (raw) -- (Decomposition) node[pos=0.5,right]{Raw Signals};
\draw[->] ($(Decomposition.south)+(.3cm,0em)$) -- ++(0,-.75) -|  ($(Reconstruction.north)+(2.5cm,0em)$) node[pos=0.35,above]{Strong  Reflectors} node[pos=0.97,left]{$\{\bold{a_m,t_m}\}$ };	
\draw[->] ($(Decomposition.south)+(-.5cm,0em)$) -- ++(0,-.75) -|  (Compression.north)  node[pos=0.35,above]{Background} node[pos=0.75,left] {$\{y_m[n]\}$};	
\path [line] (Compression) -- (Beamforming) node[pos=0.5,left]{$\{\bold{c_m}\}$ };
\path [line] (KSVD) -- (Compression) node[pos=0.5,above]{ $\bold{D}$ };
\draw[->,dashed] ($(Decomposition.south)+(-.5cm,0em)$) -- ++(0,-.75) -|  (KSVD.north);	
\path [line] (Beamforming) -- ($(Reconstruction.north)+(-2.2cm,0em)$) node[pos=0.6,left] { $\Phi[n]$ };
\node [cloud, below=0.7cm of Reconstruction] (recCombined) {Combined Image};
\node [cloud, right=0.39cm of recCombined] (recStrong) {Strong Reflectors Image};
\node [cloud, left=0.5cm of recCombined] (recBackground) {Background Image};
\path [line] (Reconstruction) -- (recCombined);
\path [line] ($(Reconstruction.south)+(-2.3cm,0em)$) -- (recBackground);
\path [line] ($(Reconstruction.south)+(+2.2cm,0em)$) -- (recStrong);
\end{tikzpicture}
}
\hfill
\caption{The proposed processing scheme. $y_m[n]$ is the  background signal obtained from the $m$-th sensor, $\bold{\{a_m,t_m\}}$ are the estimated strong reflectors parameters, $\mathbf{c_m}$ are the representation coefficients of $y_m[n]$ over the dictionary $\mathbf{D}$, and $\Phi[n]$ is the beamformed background component, composed of all sensor coefficients $\left(\{\bold{c_m}\}_{m=1}^{M}\right)$.}
\label{Fig:ProcessingScheme}
\end{figure}

The proposed processing scheme is illustrated in Figure~\ref{Fig:ProcessingScheme}.

We emphasize that the proposed algorithm is applied to the raw one-dimensional signals, as acquired by the individual sensors, immediately after their sampling. The scope of this work is therefore limited to the digital domain and does not attempt to perform compressed sensing.

Assuming each component on its own is compressible, we derive the suitable representation bases. Thereafter, the compressed background signals are integrated in a modified beamforming process, applied directly at the representation domain. To conclude the proposed algorithm, the strong reflectors are reconstructed from their sparse coefficients obtained during the decomposition stage, combined with the beamformed background signals and processed to form an image.

\subsection{Learning a Dictionary for the Background Data} \label{section:BeackgroundDictLearning}

Having separated the two components, we next want to compress them.
The strong reflectors were already compressed by saving only the pulse model parameters along the decomposition process presented in Section~\ref{section:Decomposition}.
As for the speckle (background) component, each such signal will be sparsely represented over an optimized dictionary that is trained offline from prototype examples of background signals.
It should be emphasized that this training process is only performed once for every imaging system settings, and need not be repeated for every analyzed signal or even for every imaged frame.
\newline

For the purpose of dictionary learning, we shall use the K-SVD algorithm \cite{Aharon2006}.
We chose the training set to be a subset of the signals constituting a single frame of real cardiac imaging data.
Although our goal is to compress raw signals, i.e. signals detected by each sensor prior to receive beamforming, the training set signals are chosen to be beamformed scan lines, since those were shown to have improved SNR \cite{Szabo2004}.

The input frame consists of 120 scan lines, out of which 10 scan lines were randomly chosen for the training set. Each of these signals contains 3328 samples and was divided into one-dimensional, non-overlapping patches of 100 samples each.
The dictionary size is $100\times 400$, having a redundancy factor of 4.
Initializing the dictionary with a random matrix, the training process was limited to 10 iterations.

It should be noted that the chosen training set is relatively small. 
Moreover, overlapping patches are commonly used due to the fact that the different sparse approximations could eventually be averaged at the overlapping segments, which could lower the representation error and smooth the overall signal reconstruction. Despite this apparent advantage, we chose to reduce the amount of patches (and thus the runtime of both training and representation) by using non-overlapping patches. 

However, as we show later, by training a dictionary with a small subset of \emph{beamformed} signals of a single frame, one can get a good representation for the \emph{raw echoes} of all the sensors and all the scan lines in the frame, as well as of raw signals from other frames obtained with the same system (for the same settings).

\subsection{Sparse Representation of the Background Data} 
Having the dictionary trained offline, let us return to the online processing cycle.
After decomposition, the separated background component undergoes further processing, starting with compression.

Due to computational limitations of the K-SVD algorithm, the dictionary is learned for small one-dimensional patches.
Thus each background signal is first divided into non-overlapping patches, similarly to the division of the training set signals described in the previous section.
Using the Orthogonal Matching Pursuit (OMP) algorithm, each patch is then sparsely represented over the trained dictionary.

Denote the patch length by $Q$ and the trained dictionary by $\bold{D}\in\mathbb{R}^{Q\times K}$, then using OMP we solve for each patch
\begin{equation}
\arg\underset{\mathbf{x}}{\min}\;\|\mathbf{x}\|_0 \quad \mbox{ subject to }\quad \|\mathbf{y-Dx}\|_2\leq\epsilon
\end{equation}
where $\bold{y}$ is the $Q\times 1$ extracted patch and $\bold{x}$ is its corresponding $K\times 1$ sparse coefficients vector.

Having obtained the sparse representations of all the individual patches of a given RF signal, the signal could be reconstructed.
For that matter, each patch is recovered from its sparse coefficients by $\bold{\hat{y}}=\bold{Dx}$. The recovered patches are then plugged back to the locations from which they were extracted, to reassemble the full signal.
This process is repeated for each of the raw signals detected by the different transducer elements for the same scan line. Afterwards, standard beamforming techniques can be applied to generate the beamformed background signal, which could then be further processed to form the image.
However, we propose an alternative reconstruction scheme based on performing the beamforming in the representation domain, thus constructing the beamformed signal directly from the sparse coefficients vectors of all the individual patches, rather than reconstructing each of the sensor echoes separately.
The formulation of the representation domain beamforming is presented in the next section.

\section{Beamforming in the Representation Domain} \label{section:compressedBFmain}

\subsection{Conventional Delay and Sum Beamforming}  \label{section:DAS}
\begin{figure}[h]
\centering\includegraphics[width=0.3\textwidth]{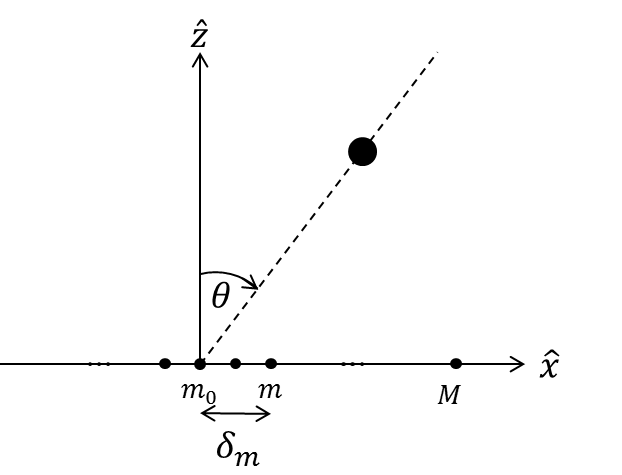}
\caption{Imaging setup} \label{Fig:setup}
\end{figure}

Modern ultrasound transducers are comprised of multiple elements. 
This inherent redundancy can be utilized by averaging the signals detected by the different sensors after aligning them with the appropriate time delays. This process, referred to as receive beamforming, allows both dynamic control over the focus of the receiving array (and thus better localization of the scattering structures) and improvement of the Signal-to-Noise Ratio (SNR) \cite{Jensen1999}. In order to achieve dynamic focusing, the focal point is swept along the beam direction, where at each coordinate the appropriate delays can be computed assuming a point reflector exists at that point. 
For simplicity, let us assume that the transducer lies along the $\hat{x}$ axis, centered at the origin ($\hat{x}=0$), and comprised of $M$ elements, as depicted in Figure~\ref{Fig:setup}.
A reference receiver denoted $m_0$ is set centered at the origin, and $\delta_m$ denotes the distance measured from the origin to the center of the $m$-th receiver.

The imaging cycle starts at time $t_0=0$, when the array transmits an acoustic pulse into the tissue at some direction $\theta$. The pulse propagates through the tissue at velocity $c$ (the speed of sound) such that at time $t\geq 0$ its coordinates are $(\hat{x},\hat{z})=(ct\sin\theta,ct\cos\theta)$. 
Consider a potential reflection, originating at this coordinate and arriving at the $m$-th element. 
The distance traveled by such a reflection is:
\begin{equation}  \label{Eq:dm}
    d_m(t,\theta) = \sqrt{(ct\cos\theta)^2+(ct\sin\theta-\delta_m)^2}
\end{equation}

The time it takes this reflection to cross that distance is $d_m(t,\theta)/c$, so it reaches the $m$-th element at time:
\begin{equation} \label{Eq:tau_m}
    \tau_m(t,\theta) = t+\frac{d_m(t,\theta)}{c}
\end{equation}

Plugging \eqref{Eq:dm} into \eqref{Eq:tau_m}:
\begin{equation} \label{Eq:tau_m_new}
\begin{aligned}
    \tau_m(t,\theta) =& t+\frac{1}{c}\sqrt{(ct\cos\theta)^2+(ct\sin\theta-\delta_m)^2} \\
			=& t+\sqrt{t^2+\left(\frac{\delta_m}{c}\right)^2-2\left(\frac{\delta_m}{c}\right)t\sin\theta}
\end{aligned}
\end{equation} 

Echoes are then reflected from density and propagation-velocity perturbations in the radiated medium, and detected by the transducer elements.
Denote by $\varphi_m(t,\theta)$ the signal detected by the $m$-th element.
In order to compensate the differences in arrival time, an appropriate delay should be applied to $\varphi_m(t,\theta)$ before these echoes could be averaged.
Let us denote the delayed signal detected by the $m$-th element by $\hat{\varphi}_m(t,\theta)$.

It is readily seen that $\tau_{m_0}(t,\theta) = 2t$.
Hence, in order to align the echo received by the $m$-th element with the one received by the reference element, we require:
\begin{equation}
    \hat{\varphi}_m(2t,\theta) = \varphi_m(\tau_m(t,\theta),\theta)
\end{equation}
Therefore the time-aligned signal received by the $m$-th element is:
\begin{equation}
    \hat{\varphi}_m(t,\theta) = \varphi_m\left(\tau_m\left(\frac{t}{2},\theta\right),\theta\right)
\end{equation}
where using \eqref{Eq:tau_m_new}:
\begin{equation} \label{Eq:timeDelay}
    \tau_m\left(\frac{t}{2},\theta\right) = \frac{t}{2}+\frac{1}{2}\sqrt{t^2+4\left(\frac{\delta_m}{c}\right)^2-4\left(\frac{\delta_m}{c}\right)t\sin\theta}
\end{equation} 

Finally, the beamformed signal is the average of the individual time-aligned echoes, i.e.
\begin{equation} \label{Eq:contBF}
    \Phi(t,\theta) = \frac{1}{M}\sum_{m=1}^{M}\varphi_m\left(\tau_m\left(\frac{t}{2},\theta\right),\theta\right)
\end{equation}

In the sequel, to simplify the notation, we will drop the dependence on 
$\theta$.

So far we have assumed continuous time signals. However, in modern ultrasound systems the receive beamforming process is performed digitally. Hence, the signals at the elements are sampled and we have $\varphi_m[n]=\varphi_m(nT)$ where $T$ is the sampling interval.
Using the sampled data for our beamforming raises a new problem, as we are now constrained to signal values on the grid $nT$. So, if the delay we need to apply is not an integer multiple of $T$, we approximate by linear interpolation
\begin{equation}
\begin{aligned}
 &\varphi_m[n] = \varphi_m\left(\tau_m\left(\frac{nT}{2}\right)\right) \\ & \quad \approx \left(1-\alpha_m[n]\right)\varphi_m[\tau_m[n]]+\alpha_m[n]\varphi_m[\tau_m[n]+1]
\end{aligned}
\end{equation}

where
\begin{equation}
\tau_m[n] = \left\lfloor\frac{\tau_m\left(\frac{nT}{2}\right)}{T}\right\rfloor
\end{equation}
and
\begin{equation}
\alpha_m[n] = \frac{\tau_m\left(\frac{nT}{2}\right)}{T}-\tau_m[n]
\end{equation}
Then, the received beamforming in its digital form is
\begin{equation} \label{Eq:BFsigDiscrete}
\Phi[n] = \frac{1}{M}\sum_{m=1}^{M}\begin{bmatrix}1-\alpha_m[n]&,&\alpha_m[n]\end{bmatrix}\left[\begin{array}{c} \varphi_m[\tau_m[n]] \\ \varphi_m[\tau_m[n]+1]\end{array}\right]
\end{equation}

\subsection{Representation Domain Beamforming} \label{section:compressedBF}
As previously discussed in Section~\ref{section:BackgroundCompression}, each sampled data vector from each element has been divided into one dimensional patches (sub-vectors) and for each we have found its sparse representation over the dictionary we learned. 

Let us denote by $\boldsymbol{\varphi}_m\in\mathbb{R}^N$ the background component of the signal received by the $m$-th sensor. Each such component is divided into $P$ patches $\bold{y}_{m,p}\in\mathbb{R}^Q$ of length $Q$:
\begin{equation} \label{Eq:phiVec}
\begin{aligned}
\boldsymbol{\varphi}_m &= \begin{bmatrix} \varphi_m[0] & \varphi_m[1] & \hdots & \varphi_m[N-1]\end{bmatrix}^T\in\mathbb{R}^N \\
&= \begin{bmatrix} \bold{y}_{m,1}^T & \bold{y}_{m,2}^T & \hdots & \bold{y}_{m,P}^T\end{bmatrix}^T
\end{aligned}
\end{equation} 
For simplicity here we assume that the patches do not overlap. However, we note that a similar derivation can be made for overlapping patches.

Let $\bold{D}\in\mathbb{R}^{Q\times K},K>Q$ denote the learned dictionary, then using OMP we solve for each patch ($\forall 1\leq m\leq M, 1\leq p\leq P$):
\begin{equation}
\arg\underset{\bold{z}_{m,p}}{\min}\|\bold{z}_{m,p}\|_0\quad \mbox{subject to}\quad\|\bold{y}_{m,p}-\bold{D}\bold{z}_{m,p}\|_2\leq\varepsilon_0
\end{equation}
where $\bold{z}_{m,p}\in\mathbb{R}^K$ but $\|\bold{z}_{m,p}\|_0<<Q$ (namely, the number of non-zero entries of all $\bold{z}_{m,p}$ is considerably smaller than the patch size). 
Following, each patch is reconstructed by $\hat{\bold{y}}_{m,p}=\bold{D}\bold{z}_{m,p}$. 
Using \eqref{Eq:phiVec} we can write
\begin{equation}
\boldsymbol{\varphi}_m = \begin{bmatrix}\bold{D} & 0 & \hdots & 0 \\ 0 & \bold{D} & \hdots & 0 \\ \vdots & \vdots & \ddots & \vdots \\ 0 & 0 & \hdots & \bold{D} \end{bmatrix} \left[\begin{array}{c} \bold{z}_{m,1} \\ \bold{z}_{m,2} \\ \vdots \\ \bold{z}_{m,P}\end{array}\right] = \widetilde{\bold{D}}\bold{z}_m
\end{equation}
where $\bold{\widetilde{D}}\triangleq \bold{I}_P\otimes \bold{D}$  and $\otimes$ denotes the Kronecker tensor product, namely $\bold{A}\otimes\bold{B}=[a_{ij}\bold{B}]_{i,j}$ for matrices $\bold{A}$ and $\bold{B}$.

Substituting in \eqref{Eq:BFsigDiscrete} we obtain for the beamformed signal
\begin{equation}
\begin{aligned}
\Phi[n] &= \frac{1}{M}\sum_{m=1}^{M}\begin{bmatrix}1-\alpha_m[n]&\alpha_m[n]\end{bmatrix}\left[\begin{array}{c} \bold{e}_{\tau_m[n]}^T \\ \bold{e}_{\tau_m[n]+1}^T \end{array}\right]\widetilde{\bold{D}}\bold{z}_m \\
&= \frac{1}{M}\sum_{m=1}^{M} \bold{b}_m[n]^T\bold{z}_m
\end{aligned}
\end{equation}
where $\bold{e}_l$ is the $l$-th column of the $N\times N$ identity matrix $\bold{I}_N$ and $\bold{b}_m[n]\in\mathbb{R}^{KP}$, and we recall that $\tau_m[n]$ are indices from $(1,2,...,N)$.

Finally, denote:
\begin{equation}
\bold{B}_m = \begin{bmatrix}
\bold{b}_m[1] & \bold{b}_m[2] & \hdots & \bold{b}_m[N]
\end{bmatrix}^T \in\mathbb{R}^{N\times PK}
\end{equation}

\begin{equation}
\bold{H} = \frac{1}{M}\begin{bmatrix}
\bold{B}_1 & \bold{B}_2 & \hdots & \bold{B}_M
\end{bmatrix}
\end{equation}

\begin{equation}
\bold{Z} = \begin{bmatrix}
\bold{z}_1^T & \bold{z}_2^T & \hdots & \bold{z}_M^T
\end{bmatrix}^T
\end{equation}

Then the beamformed signal is given by
\begin{equation}
\bold{\Phi} = \bold{H}\bold{Z}
\end{equation}

It is readily seen that $\bold{H}$ depends solely on the imaging settings and the learned dictionary.
This result indicates that beamforming in the representation domain is manifested as a weighted combination of the representation coefficients, where the weight matrix $\bold{H}$ is data independent and can be computed apriori for each scan line direction $\theta$.
Therefore, only the sparse coefficients should be transferred to the beamformer, leading to a much simplified computation of the beamformed signal.

\subsection{Strong Reflectors Reconstruction} 
So far we discussed the processing and reconstruction of the background component. 
The strong reflectors may be reconstructed from their sparse coefficients obtained during the decomposition stage, and combined with the beamformed background signals to form an image, without going through a beamforming process.
 
Assuming a known pulse shape, the strong reflectors reconstruction is straightforward. The estimated coefficients could be directly plugged into the parametric pulse model, thus recovering the time samples of the strong reflectors component. These recovered signals could then undergo the standard beamforming and image formation stages.
Nonetheless, this reconstruction process is unnecessary, and as point reflectors are concerned, their location could be derived from the measurements of all the elements.
Having an estimated time-delay in each element, this delay defines a circle of possible locations whose radius is the reflection's distance from that element.
The spatial location of the strong reflector can therefore be extracted from the intersection of each two such circles. 
All in all, the strong reflectors could be reconstructed directly into the beamformed signal, saving the need for an additional beamforming process.
\newline

\section{Side Lobes Suppression} \label{section:SideLobeSuppression}
\label{section:modifiedSeparation}
One of the imaging artifacts resulting from the receive beamforming process is the side lobes artifact.
Though aspiring to transmit a narrow beam, the actual transmission is not solely concentrated in the main axis, but rather radiates a wide region according to a sinc-shaped beam-profile, such that the beam has maximal energy on the axis but some of the energy is dispersed in side lobes.

The beamforming process concentrates the received beam by assuming that all the received reflections originated from the main axis. Consequently, echoes from a scatterer in the side lobe pathway are erroneously perceived coming from the main beam, resulting in image artifacts, manifested as smeared out echoes. These artifacts not only degrade the contrast of the image and impair its visual quality, but may also compromise its diagnostic credibility \cite{Szabo2004}.
Side lobes suppression is usually treated by apodization, which degrades the lateral resolution, or by adaptive beamforming \cite{Vignon2008}.
\newline

As the side lobe artifacts result from the presence of strong reflectors during the beamforming process, we could utilize the decomposition approach so as to perform beamforming only for the background component, after separating the strong reflectors.

When evaluating the ability of our decomposition methods to reduce side lobes artifacts, we note that the success of these methods depends on the gain of the side lobe reflections with respect to its surroundings.
Recall that both proposed decomposition methods are based on an amplitude threshold as a stopping criterion. Therefore, if the side lobe reflection is weak compared with speckle reflections along the same beam, no amplitude threshold will enable its detection as a strong reflector.

In order to account for such cases as well, we propose the following alteration to our decomposition methods.
A greedy search similar to the one previously described is applied to each of the raw signals. In each iteration, the best match for the strong pulse that is above the amplitude threshold is detected and subtracted using either the STFT of IQ based method. 
Unlike the previous methods, before proceeding to the next iteration, the time-delay of the detected pulse is here adjusted for all the other sensors and for the adjacent scan lines. 
For these other signals, a pulse will be matched and subtracted if a local maxima occurs at the adjusted expected time, regardless of the predefined threshold.
The time adjustment is calculated similarly to the delay computed for the purpose of receive beamforming (see Section~\ref{section:DAS}).

Explicitly, if a point reflector is placed at distance $r$ and direction $\theta$ from the array center (see Figure~\ref{Fig:setup}), then its distance from the $m$-th sensor, positioned along $\hat{z}=0$ at $\hat{x}=\delta_m$, is:   
\begin{equation} 
r_m=\sqrt{\left(r\cos\theta\right)^2+\left(r\sin\theta-\delta_m\right)^2}
\end{equation}
and the echo reflected from it arrives to the $m$-th sensor at time $t_m=t_{TX}+\frac{r_m}{c}$, where $t_{TX}=\frac{r}{c}$.
From $t_m$ we can extract:
\begin{equation}
r=\frac{c^2t_m^2-\delta_m^2}{2(ct_m-\delta_m\sin\theta)}
\end{equation}
Then, the expected time of arrival to the $k$-th sensor is:
\begin{equation} \label{Eq:timeToSensorK}
t_k=t_{TX}+\frac{r_k}{c}=\frac{r}{c}+\frac{1}{c}\sqrt{\left(r\cos\theta\right)^2+\left(r\sin\theta-\delta_k\right)^2}
\end{equation}

The modification of the decomposition algorithm as applied to the $l$-th scan line is summarized in Algorithm~\ref{Alg:modifiedDecomposition}.
\begin{algorithm}[ht]
\caption{Modified Decomposition Algorithm for the $l$-th scan line (example given for the STFT-based method)}
\label{Alg:modifiedDecomposition}
\begin{algorithmic}[]
\STATE for each sensor $m=1,...,M$
\begin{itemize}
\item run Algorithm~\ref{Alg:STFTdecomposition} 
and obtain the background $\bold{r^k}$ and strong reflectors parameters $\{t_j,\omega_j,a_j\}_{j=1}^{k}$.
\item for each identified pulse, adjust its time delay to all sensors and to adjacent scan lines:

for pulse $j=1,...,k$

\quad for line $q=l-1,l,l+1$

\qquad for sensor $s=1,...,M$

\begin{itemize}
 \item compute $t_s^j$ using \eqref{Eq:timeToSensorK}
\item if $t_s^j$ is a local maxima for line $q$ and sensor $s$, with amplitude $a_{q,s}$, subtract $a_{q,s}\bold{H}(t-t_s,\omega-\omega_j)$ from the relevant residual
\end{itemize}
\end{itemize}
\end{algorithmic}
\end{algorithm}

The modified method here presented has a significant impact in cases where the strong reflectors are placed in a speckled region and have visible side lobe artifacts, such as the synthetic cyst phantom example described in Section~\ref{section:cystResults}.

\section{Simulation and Results} \label{section:SimulationResults}

We have conducted experiments using simulated data and real cardiac data.
In the following, we describe in some detail the setup and information pertaining to these experiments.

\subsection{1D Simulation Results} \label{section:1Dsim}

To evaluate the decomposition algorithms, both the STFT and IQ based methods for removing the strong reflectors were first tested on a single scan line simulated using the FieldII simulation program \cite{Jensen1996}.

We created an aperture comprising 64 transducer elements, with central frequency $f_0=3.5MHz$. 
The width of each element, measured along the $\hat{x}$ axis, was $\frac{1}{2}\lambda=\frac{c}{2f_0}=0.22 mm$, and its height, measured along the $\hat{y}$ axis, was $5 mm$.
The elements were arranged along the $\hat{x}$ axis, with a $0.055 mm$ inter-element spacing (kerf).
The transmitted pulse was simulated by exciting each element with two periods of a sinusoid of frequency $f_0$, where the delays were adjusted such that the transmission focal point was at depth $r=70 mm$.
No apodization was used.

The simulation parameter settings are summarized in Table \ref{tab:fieldParams1D}.

\begin{figure}[h]
\centering\includegraphics[width=0.5\textwidth]{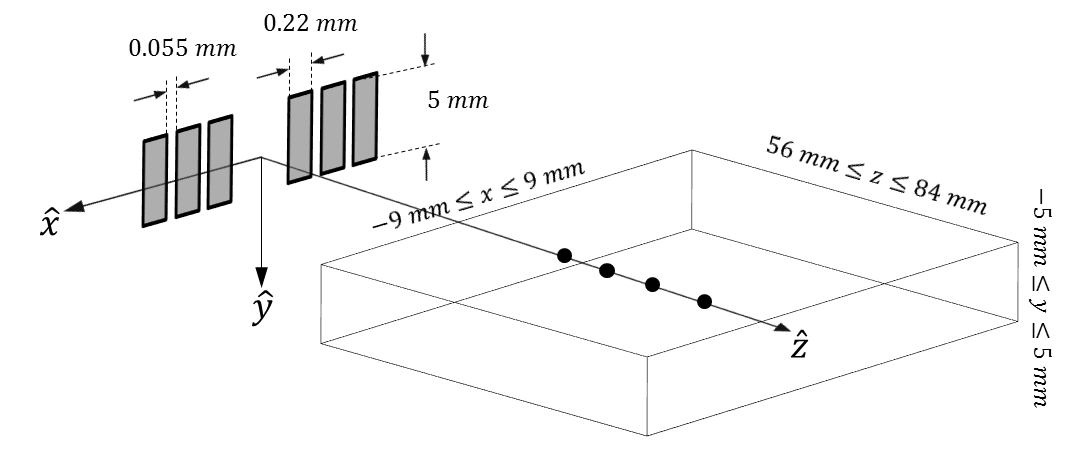}
\caption{FieldII simulation setup for the point reflectors phantom} \label{Fig:fieldSetup1D}
\end{figure}

The simulation setup is illustrated in Figure~\ref{Fig:fieldSetup1D}.

\begin {table}[ht]
\centering
\begin{tabular}{|c||c|} 
\hline 
  Parameter  & Value \\ 
\hline \hline
$c$ (Speed of Sound) & 1540 m/sec \\
\hline 
$f_0$ (Central Frequency) & 3.5 MHz \\
\hline 
$f_s$ (Sampling Frequency) & 16 MHz \\
\hline 
Element Width & 0.22 mm \\
\hline 
Element Height & 5 mm \\
\hline 
Kerf & 0.055 mm \\
\hline 
Number of Elements & 64 \\
\hline 
Apodization & none \\
\hline
TX Focus & 70 mm\\
\hline
\end{tabular} 
\caption{FieldII simulation parameters for the point reflectors phantom} \label{tab:fieldParams1D}
\end{table}

A single scan line was simulated for a steering angle $\theta=0^{\circ}$.

The simulated phantom consists of speckled background with 4 strong reflectors positioned $5 mm$ apart along the $\hat{z}$ axis ($\theta=0^{\circ}$).

The speckled region was constructed by randomly drawing $10^5$ point reflectors distributed uniformly in the three dimensional region given by
$\left\{\left(x,y,z\right):|x|\leq 9 mm,\;|y|\leq 5 mm,\;|z-70|\leq 14 mm\right\}$.
The corresponding amplitudes were also drawn at random according to a Normal distribution with zero-mean and unit-variance.
Then, the strong reflectors were added along the $\hat{z}$ axis at depths $65mm,70mm,75mm,80mm$.
The strong reflectors amplitudes were set to be 50 times the variance of the speckle reflectors amplitudes.
\newline

\begin{figure*}[htp]
\centering \subfloat[]{
	\centering \includegraphics[trim=0.6cm 0cm 1.1cm 0.2cm,clip,scale=0.3]{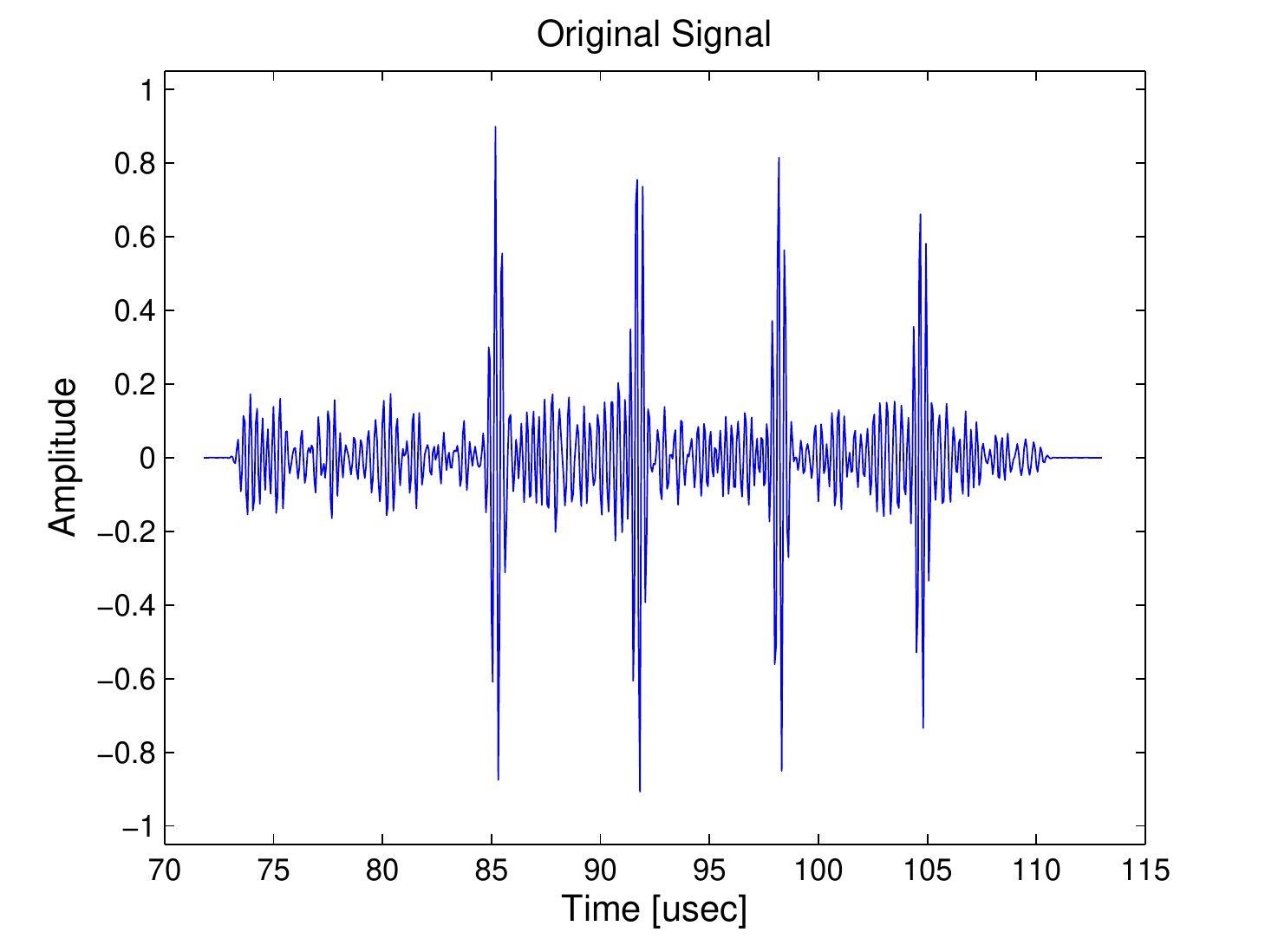}
    \label{Fig:4reflectors_orig}
} \\
\begin{tabular}{cc}
\subfloat[]{
	\centering \includegraphics[trim=0.6cm 0cm 1.1cm 0.2cm,clip,scale=0.3]{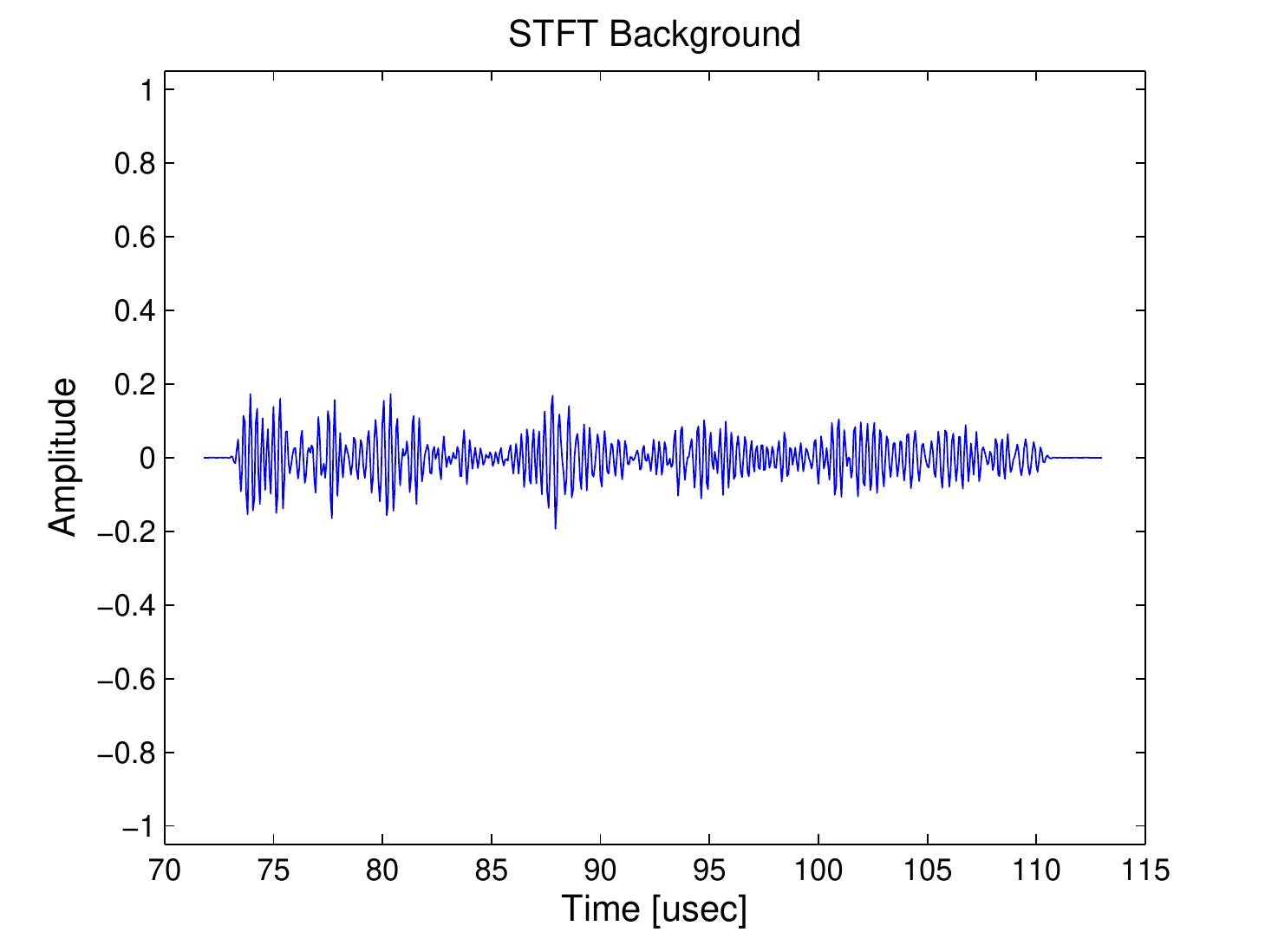}
    \label{Fig:4reflectors_stft_back}
} \qquad & \qquad 
\subfloat[]{
	\centering \includegraphics[trim=0.6cm 0cm 1.1cm 0.2cm,clip,scale=0.3]{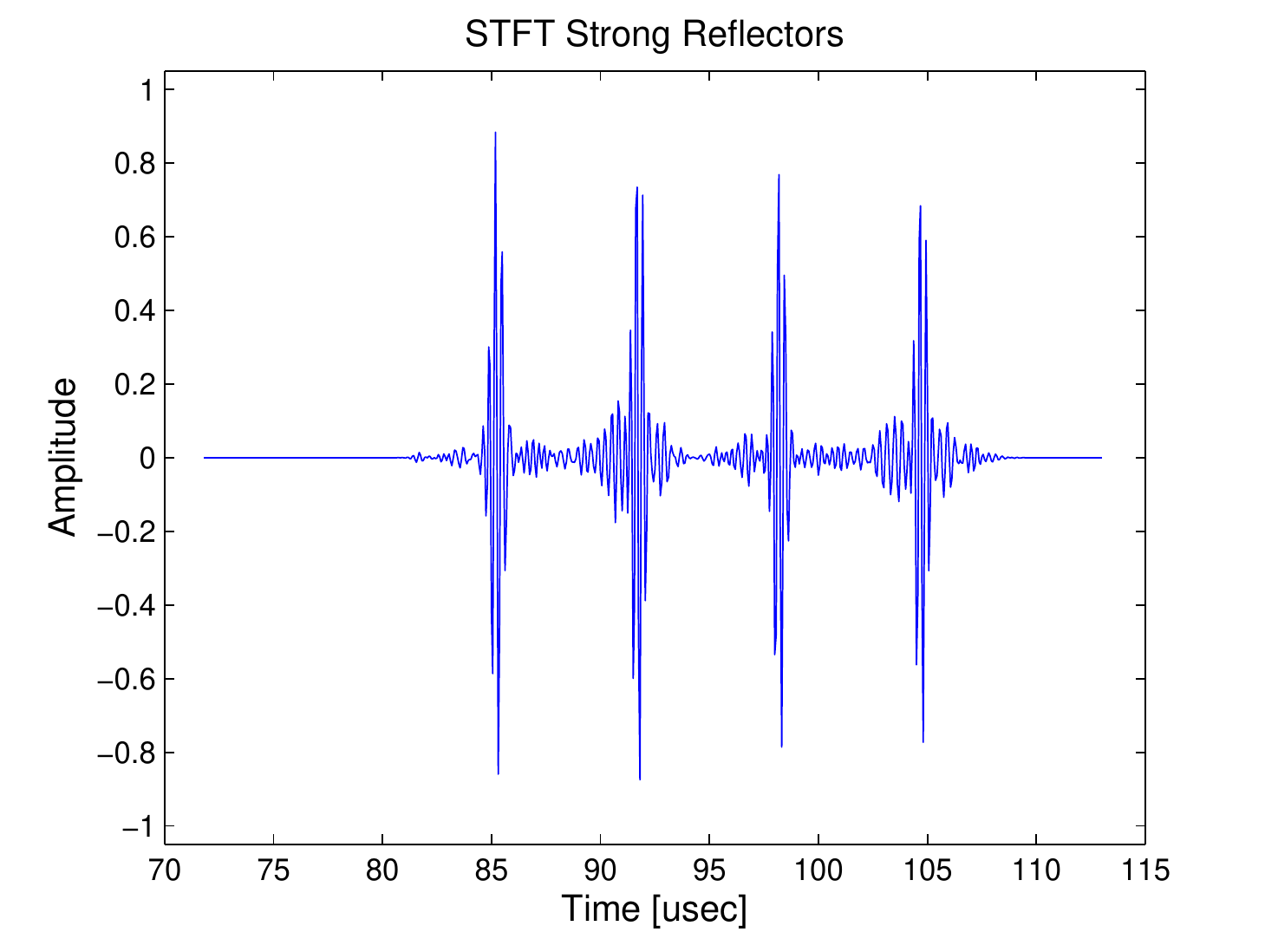}
    \label{Fig:4reflectors_stft_pulse}
}
\end{tabular}
\\
\begin{tabular}{cc}
\subfloat[]{
	\centering \includegraphics[trim=0.6cm 0cm 1.1cm 0.2cm,clip,scale=0.3]{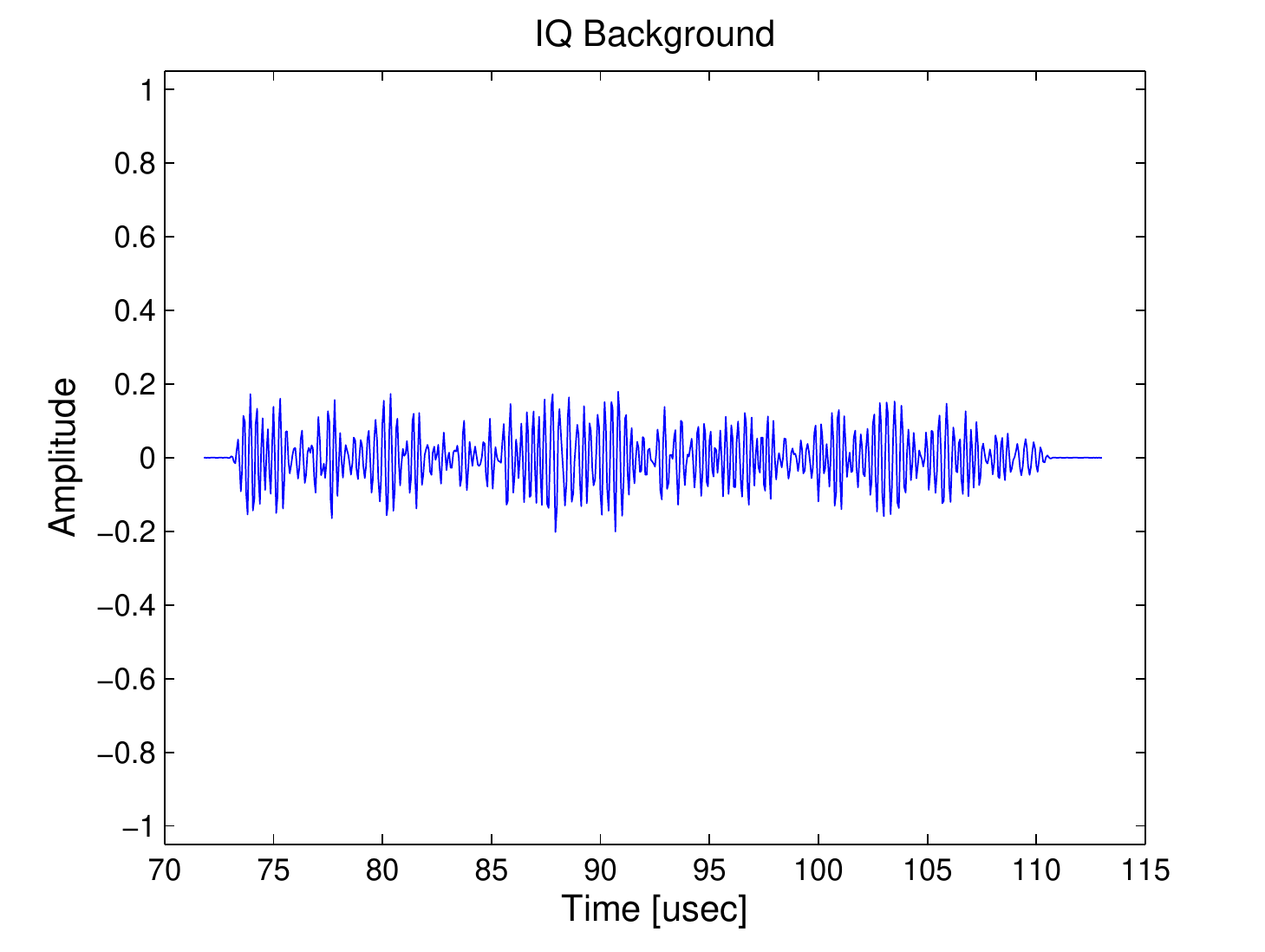}
    \label{Fig:4reflectors_iq_back}
} \qquad & \qquad 
\subfloat[]{
	\centering \includegraphics[trim=0.6cm 0cm 1.1cm 0.2cm,clip,scale=0.3]{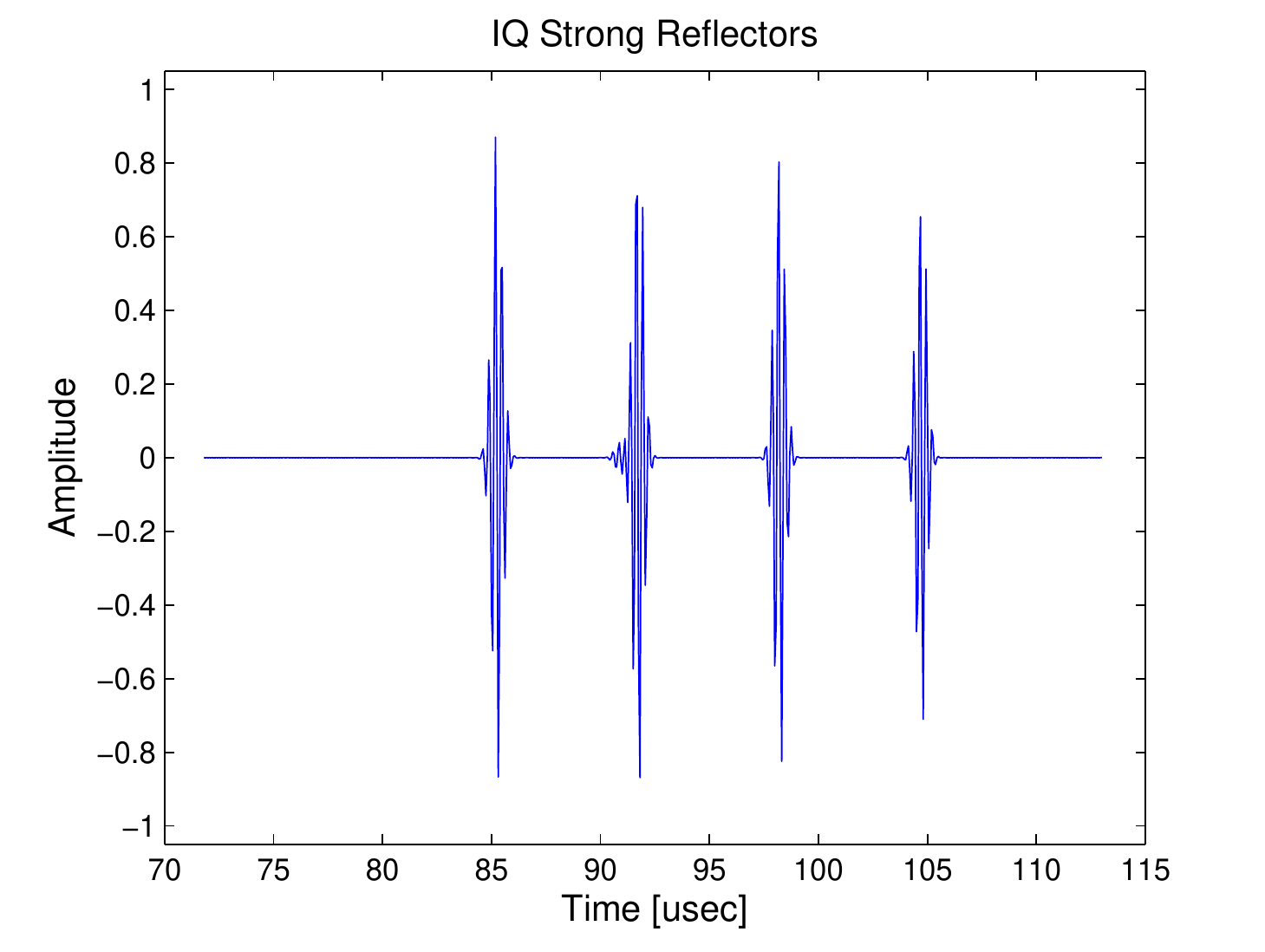}
    \label{Fig:4reflectors_iq_pulse}
}
\end{tabular}
\\
\begin{tabular}{cc}
\subfloat[]{
	\centering \includegraphics[trim=0.6cm 0cm 1.1cm 0.2cm,clip,scale=0.3]{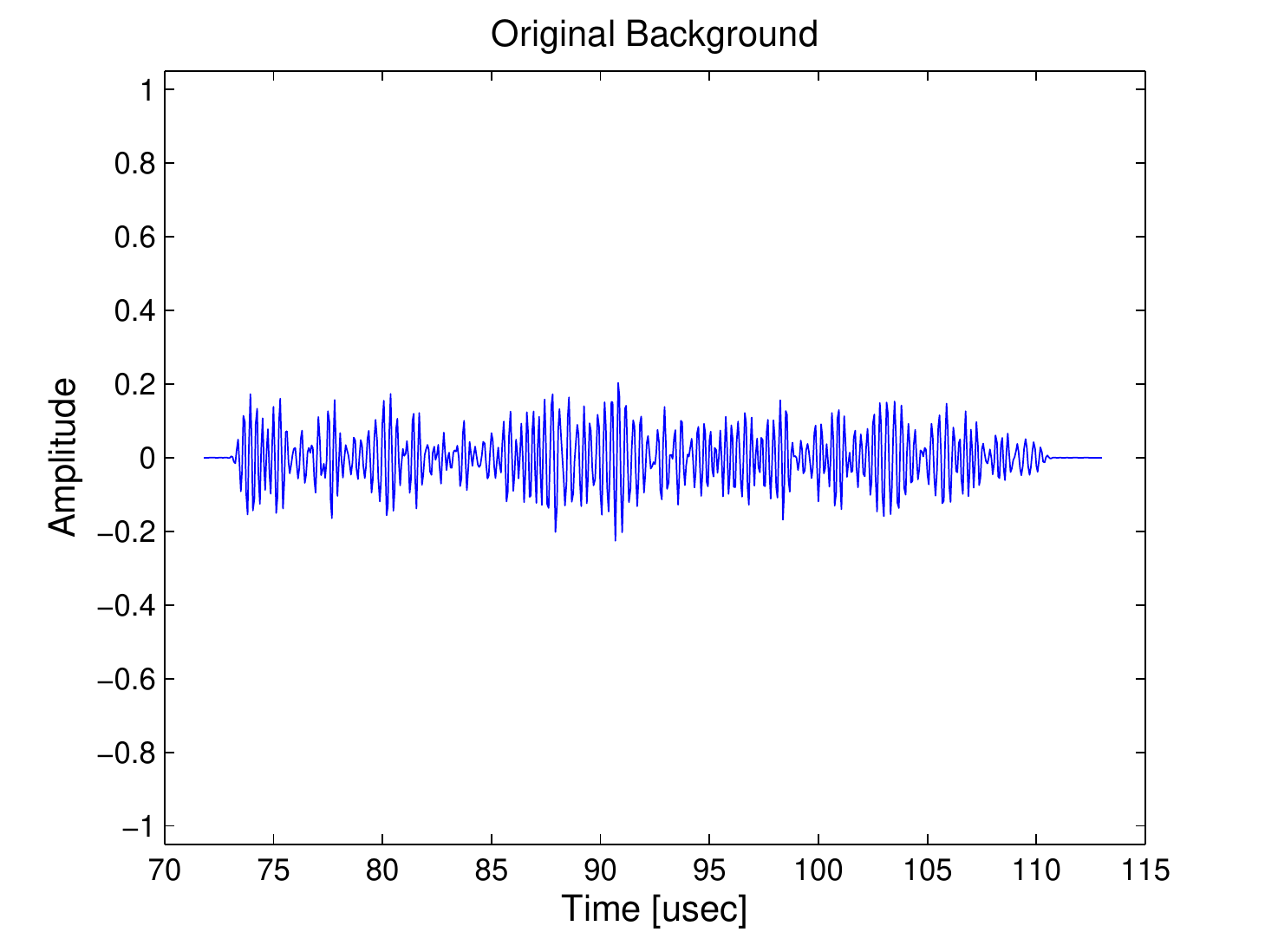}
    \label{Fig:4reflectors_gt_back}
} \qquad & \qquad 
\subfloat[]{
	\centering \includegraphics[trim=0.6cm 0cm 1.1cm 0.2cm,clip,scale=0.3]{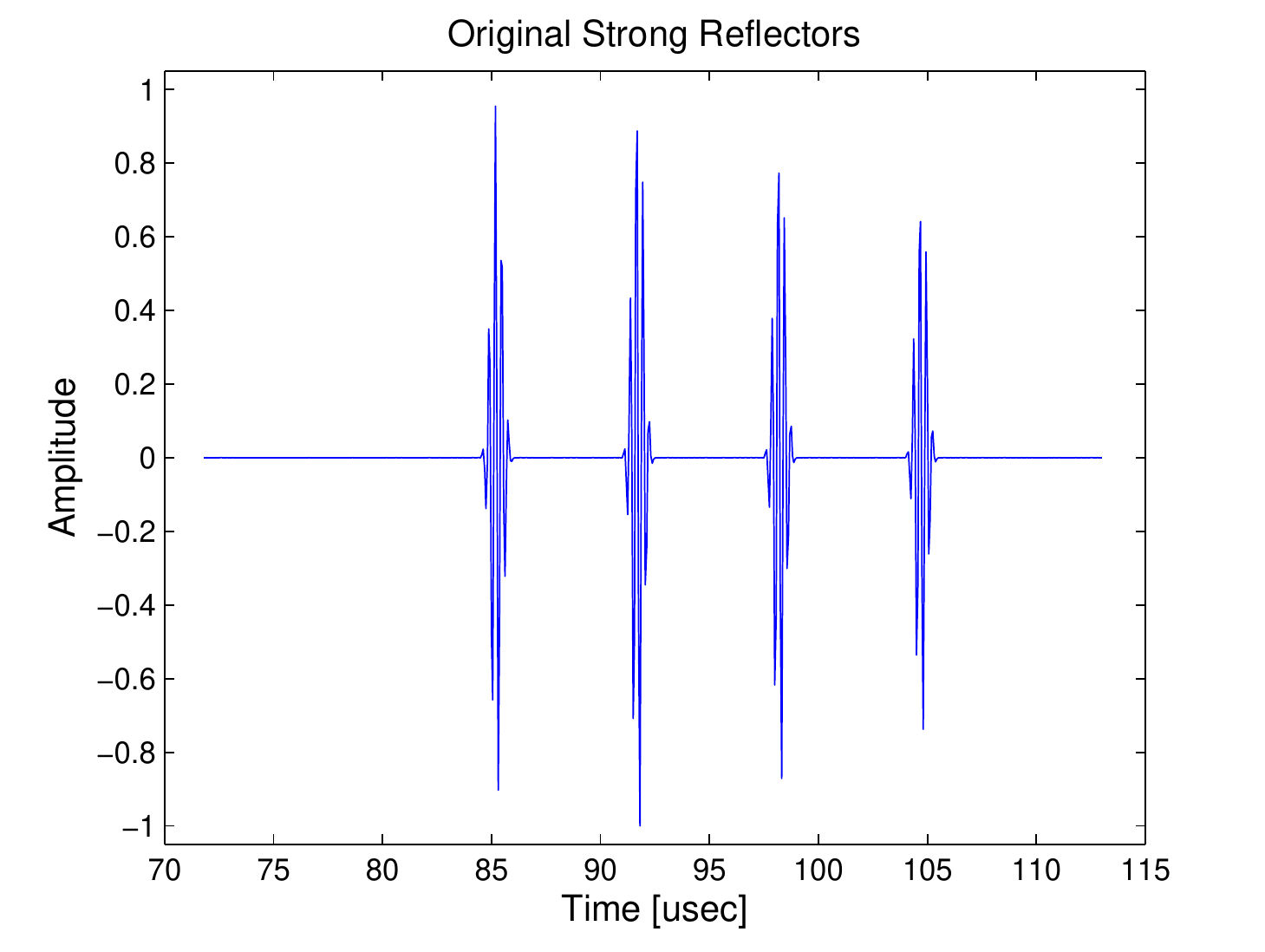}
    \label{Fig:4reflectors_gt_pulse}
}
\end{tabular}
\caption[]{Point reflectors RF signal with its decomposition: \subref{Fig:4reflectors_orig} Received RF signal, \subref{Fig:4reflectors_stft_back} Background component (STFT decomposition), \subref{Fig:4reflectors_stft_pulse} Strong reflectors (STFT decomposition), \subref{Fig:4reflectors_iq_back} Background component (IQ decomposition), \subref{Fig:4reflectors_iq_pulse} Strong reflectors (IQ decomposition), \subref{Fig:4reflectors_gt_back} Background component (ground truth), \subref{Fig:4reflectors_gt_pulse} Strong reflectors (ground truth)}
\label{Fig:4reflectors_results}
\end{figure*}

Figure~\ref{Fig:4reflectors_results} depicts the RF echo received by one of the transducer elements with its decomposition results using both STFT and IQ approaches.

These decomposition results are compared with a ground truth obtained by individual simulation of the two components.

Comparing the two approaches, it appears that the IQ based method gives a better separation of the two components, where less of the background component is removed with the strong reflectors.
The visual superiority is also reflected by the Mean Square Error (MSE) and Peak Signal-to-Noise Ratio (PSNR) calculated with respect to the ground truth components, as summarized in Table~\ref{tab:MSEresult}.

This result is consistent with our formerly mentioned expectations, since using lower frequencies and relieving the need for phase optimization makes the estimation less bound to errors.

Recall that PSNR is defined as $PSNR = 10\log_{10}{\left(\frac{I_{max}^2}{MSE}\right)}$
where $I_{max}$ is the maximal possible value of the compared signals or images (for images spanning the full 8-bit gray-scale, $I_{max}=255$) and $MSE=\mathbb{E}\left[\left(\bold{I}-\bold{\hat{I}}\right)^2\right]$ with $\bold{I}$ and $\bold{\hat{I}}$ being the compared signals or images.

\begin {table}[htp]
\centering
\begin{tabular}{|c||c|c|c|} 
\hline 
  & STFT & IQ & \begin{tabular}{c} Original \\ Signal  \ref{Fig:4reflectors_orig} \end{tabular} \\ 
\hline \hline
PSNR & 30.231 & 34.204 & 18.714 \\
\hline 
MSE & 3.79e-40 & 1.52e-40 & 5.37e-39 \\
\hline 
\end{tabular} 
\caption[PSNR and MSE results]{PSNR (in [dB]) and MSE results for the point reflectors phantom. Results are relative to the ground truth components \ref{Fig:4reflectors_gt_back}-\ref{Fig:4reflectors_gt_pulse}} \label{tab:MSEresult}
\end{table}

\subsection{Cyst Phantom Results} \label{section:cystResults}
Our four proposed methods for identifying and removing the strong reflectors (STFT, IQ, modified STFT and modified IQ) have been tested on a synthetic phantom simulated using the FieldII program \cite{Jensen1996}.
The data acquisition setup is similar to the previous phantom as summarized in Table~\ref{tab:fieldParams1D}.

During this simulation, a $24^{\circ}$ sector was imaged using 48 scan lines in Single Line Acquisition mode (SLA), i.e. a single reception line was computed for each transmission.

The phantom comprises of a large cyst in a speckled background, with a single strong reflector placed in the speckled region right besides the cyst.

The speckled region was constructed by randomly drawing $10^5$ point reflectors distributed uniformly in the three dimensional region given by
$\left\{\left(x,y,z\right):|x|\leq 9 mm,\;|y|\leq 5 mm,\;|z-70|\leq 14 mm\right\}$.
The corresponding amplitudes were also drawn at random according to a Normal distribution with zero-mean and unit-variance.
The cyst was then created by removing all point reflectors from a circle of radius $8.5mm$ centered at $\left(x,y,z\right) =  \left(0,0,70mm\right)$.
Finally, a single strong reflector was added at $\left(x,y,z\right) =  \left(8.6mm,0,70mm\right)$, i.e. at a depth of $r=70.5 mm$ at an angle of $7^{\circ}$, such that it lies exactly along the main direction of the 39-th scan line.
The strong reflector's amplitude was set to be 100 times the variance of the speckle reflectors amplitudes.
\newline

The reflector was intentionally placed very close to the cyst ($0.1mm$ from its boundary). 
While the beam focus is aimed at the cyst, side lobes might insonify the strong reflector that is positioned outside it, and its reflection would be interpreted as if it originated from within the cyst. This causes undesired artifacts in the resulting image, as previously mentioned in Section~\ref{section:modifiedSeparation}.
Had the strong reflector been placed in a speckled region, its reflections might have blended with other weak reflections in its speckled surroundings, but in our example, due to the low echogenicity of the cyst, the side lobe reflections are expected to be easily observed. 

Using this phantom, we would like to show that our methods, operating at the sensor level, are able to remove the side lobe reflections, such that these artifacts no longer exist, or are significantly decreased, in the output image.

A ground truth image was generated by repeating the simulation without the strong reflector, thus obtaining a clean background image.
\newline

\begin{figure*}[!htp]
\centering
\begin{tabular}{ccc}
\begin{tabular}{c}
\subfloat[]{
	\centering \includegraphics[trim=4.5cm 1.1cm 3.5cm 1.1cm,clip,scale=0.45]{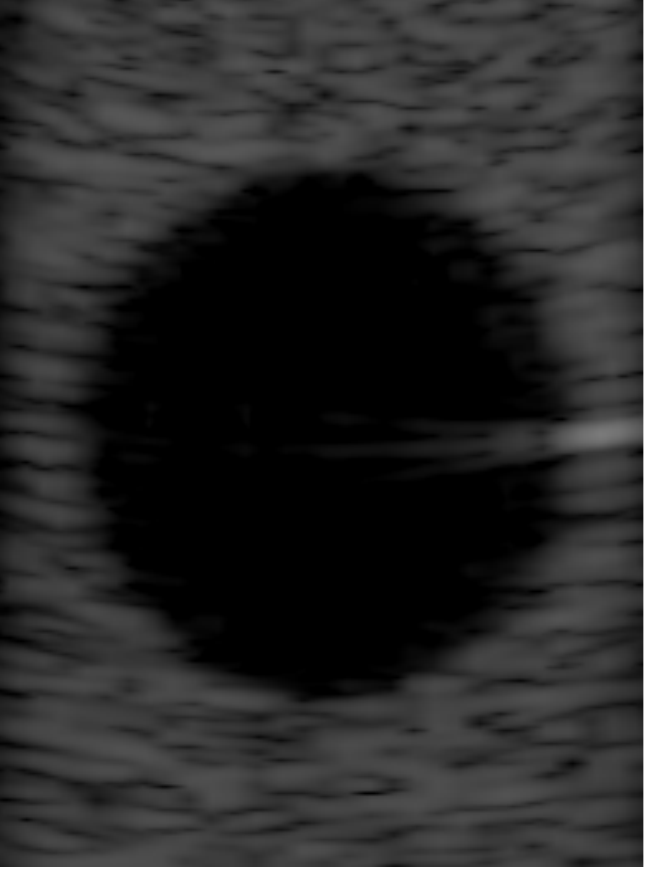}
    \label{Fig:cyst1}
}\\
\subfloat[]{
	\centering \includegraphics[trim=4.5cm 1.1cm 3.5cm 1.1cm,clip,scale=0.45]{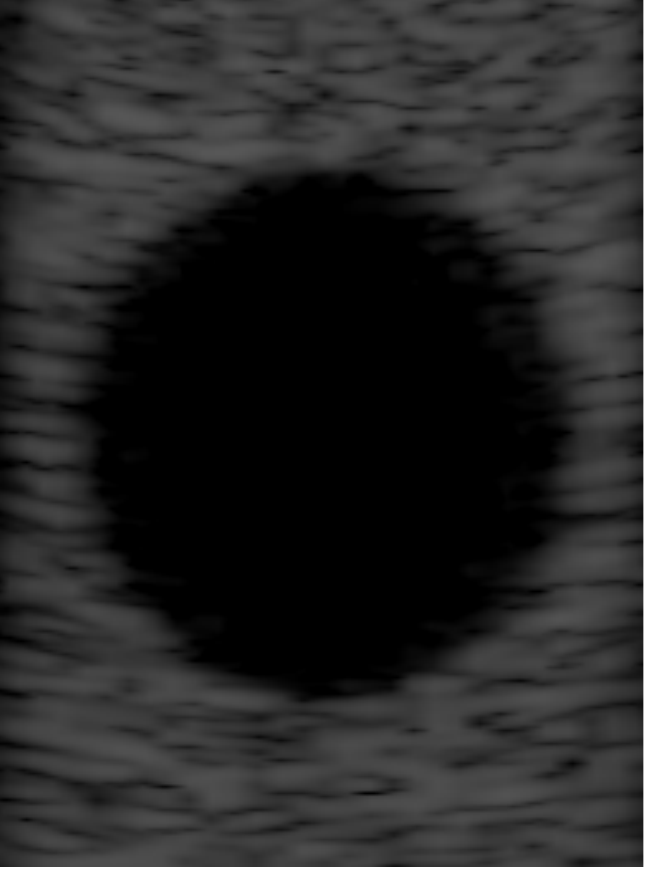}
    \label{Fig:cyst2}
}
\end{tabular} &
\begin{tabular}{c}
\subfloat[]{
	\centering \includegraphics[trim=4.5cm 1.1cm 3.5cm 1.1cm,clip,scale=0.45]{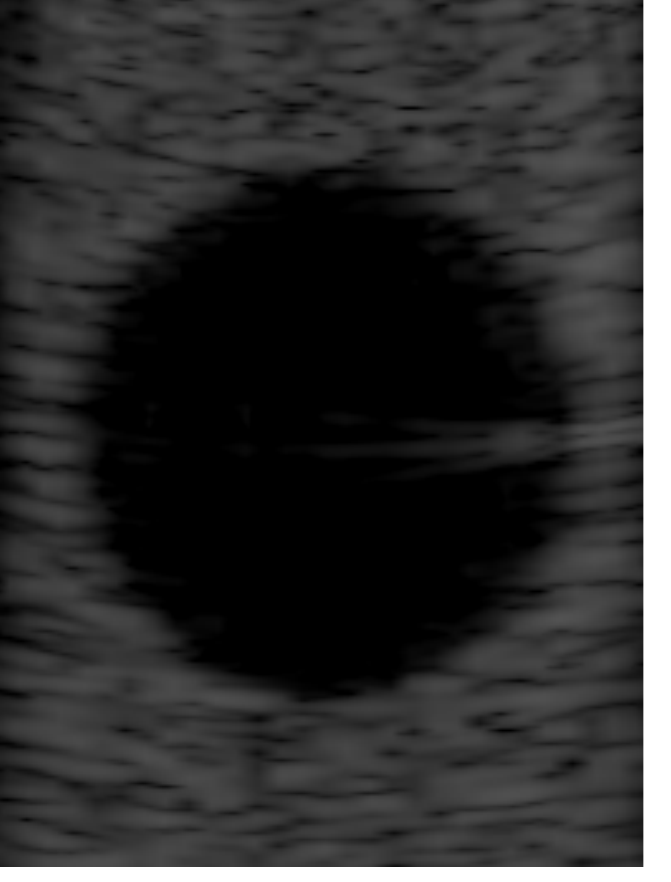}
    \label{Fig:cyst3}
}\\
\subfloat[]{
	\centering \includegraphics[trim=4.5cm 1.1cm 3.5cm 1.1cm,clip,scale=0.45]{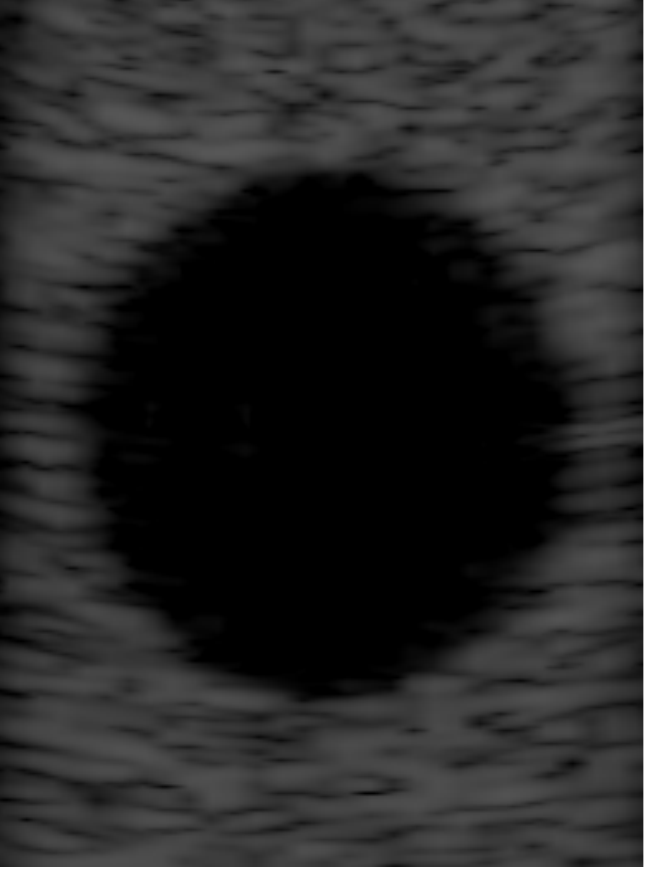}
    \label{Fig:cyst4}
}
\end{tabular} &
\begin{tabular}{c}
\subfloat[]{
	\centering \includegraphics[trim=4.5cm 1.1cm 3.5cm 1.1cm,clip,scale=0.45]{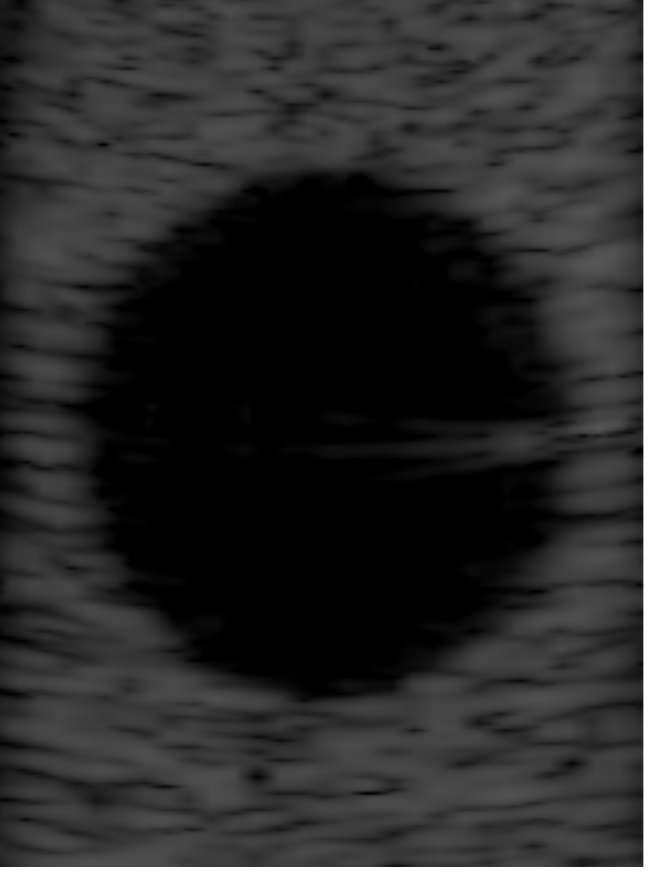}
    \label{Fig:cyst5} 
}\\
\subfloat[]{
	\centering \includegraphics[trim=4.5cm 1.1cm 3.5cm 1.1cm,clip,scale=0.45]{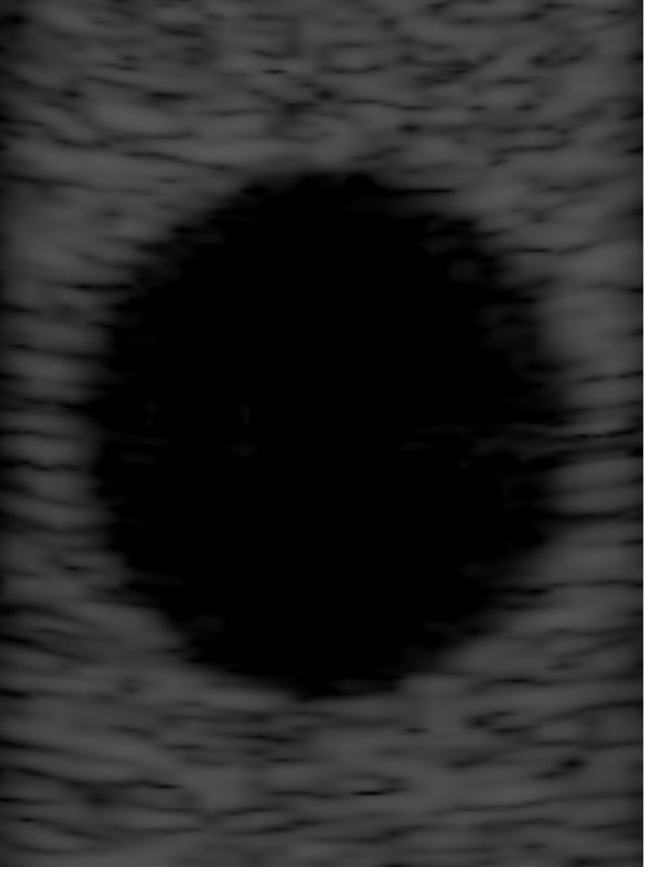}
    \label{Fig:cyst6}
}
\end{tabular}
\end{tabular}
\caption[]{Cyst phantom background estimation results: \subref{Fig:cyst1} Original image, \subref{Fig:cyst2} Original image without strong reflectors (ground truth), \subref{Fig:cyst3} STFT background image, \subref{Fig:cyst4} Modified STFT background image, \subref{Fig:cyst5} IQ background image, \subref{Fig:cyst6} Modified IQ background image.}
\label{Fig:CystResults}
\end{figure*}

The background estimation results obtained for the cyst phantom are depicted in Figure~\ref{Fig:CystResults}.
Corresponding PSNR and SSIM \cite{SSIM} values compared with the ground truth image are presented in Table~\ref{tab:cystPSNR}.

\begin{figure}[!htp]
\centering
\subfloat[]{
    \centering \includegraphics[trim=4.5cm 1cm 3.5cm 1cm,clip,scale=0.4]{cyst1a} \label{Fig:cystOrig}
}
\subfloat[]{
    \centering \includegraphics[trim=4.5cm 1cm 3.5cm 1cm,clip,scale=0.4]{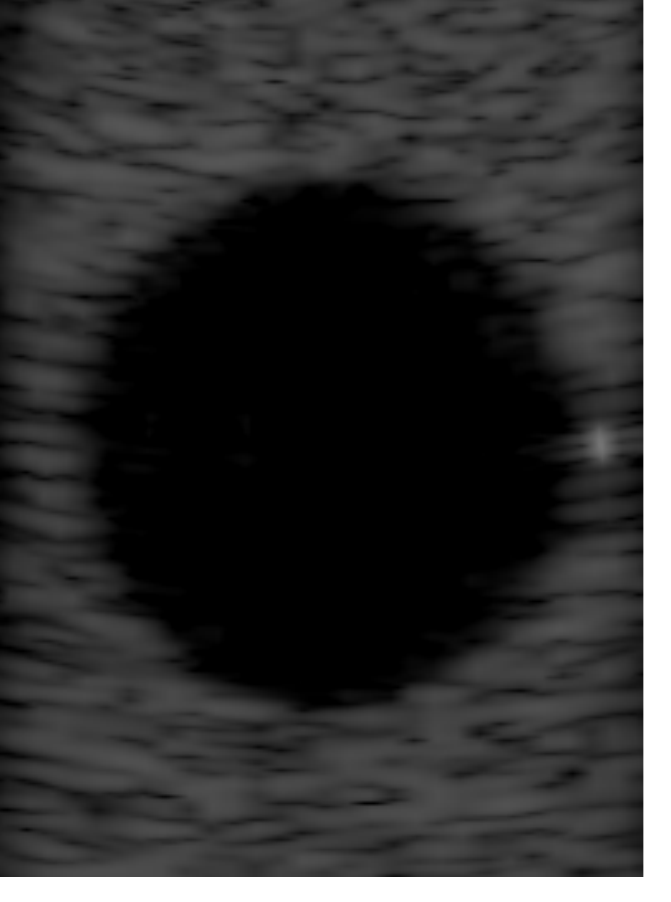}  \label{Fig:cystCombinedSTFT}
}
\subfloat[]{
        \centering \includegraphics[decodearray={0.1 0.12 0.1 0.12 0.1 0.12 0.1 0.12},trim=4.5cm 1cm 3.5cm 1cm,clip,scale=0.4]{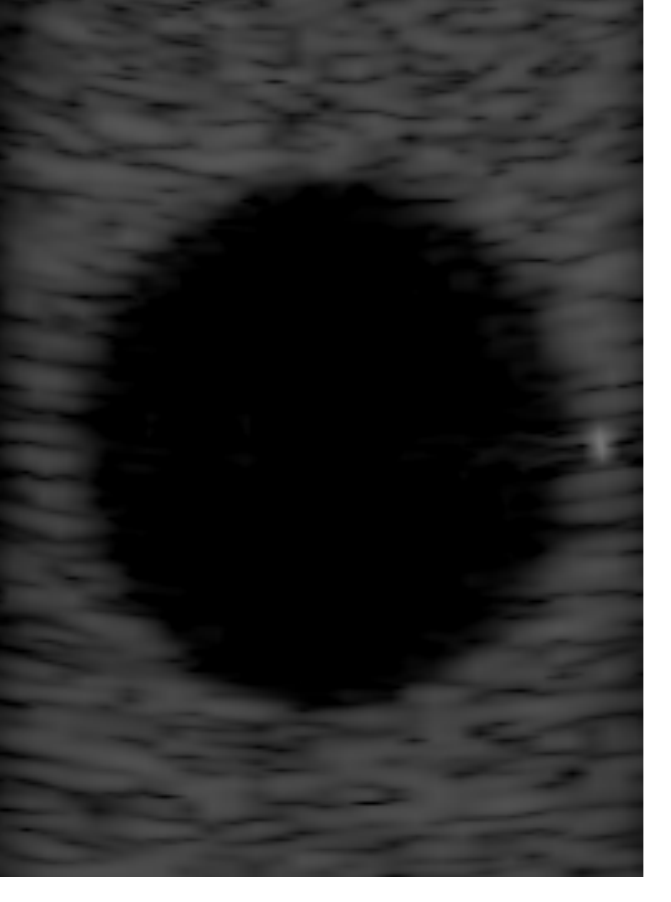}  \label{Fig:cystCombinedIQ}

}
\caption[Cyst phantom combined image results] {Cyst phantom combined image results: \subref{Fig:cyst1} Original image, \subref{Fig:cystCombinedSTFT}  Combined image using modified STFT decomposition, \subref{Fig:cystCombinedIQ} Combined image using modified IQ decomposition.}
\label{Fig:CystResultsCombined}
\end{figure}

The reconstructed combined images, obtained after adding the estimated strong reflectors to the beamformed background signals, are depicted in Figure~\ref{Fig:CystResultsCombined}.
These results show that both our modified decomposition methods (based on either STFT or IQ) yield good quality images, in which the smearing of the strong reflector due to side lobes artifacts is removed and replaced with the actual point reflector, such that the cyst boundary is no longer obscured.

\begin {table}[htp]
\scalebox{0.96}{
\begin{tabular}{|c||c|c|c|c|c|} 
\hline 
  & \begin{tabular}{c} Orig. \\ Image \\ \ref{Fig:cyst1} \end{tabular} & \begin{tabular}{c} STFT \\ \ref{Fig:cyst3} \end{tabular} & \begin{tabular}{c} Modified \\ STFT \\ \ref{Fig:cyst4} \end{tabular} & \begin{tabular}{c} IQ \\ \ref{Fig:cyst5} \end{tabular} & \begin{tabular}{c} Modified \\ IQ \\ \ref{Fig:cyst6} \end{tabular} \\ 
\hline \hline
PSNR & 34.56 & 33.28 & 40.08 & 36.03 & 43.27 \\ 
\hline 
SSIM & 0.9925 & 0.9848 & 0.9947 & 0.9901 & 0.9957 \\ 
\hline 
\end{tabular} 
}
\caption[PSNR and SSIM results for the cyst phantom]{PSNR (in [dB]) and SSIM results for the cyst phantom. Results are relative to the ground truth image \ref{Fig:cyst2}} \label{tab:cystPSNR}
\end{table}

Based on these results, the modified decomposition method presented in Section~\ref{section:modifiedSeparation} therefore seems better in terms of removing side lobes. Nevertheless, this modification has its down-side. Now depending on adjacent lines, this method can no longer run in parallel for all the sensor echoes of all the scan lines, but should rather run on a collection of such lines, thus dictating a longer runtime compared with the independent removal methods. 

It should be emphasized that this simulated example represents the worst-case scenario.
In some cases, a suitable amplitude threshold, i.e. one that separates the side lobe reflections from the background, may exist for the non-modified methods as well.
Additionally, in more typical cases the strong reflectors are placed within echogenic tissues, such that the side lobe reflections themselves, or any minor errors in their detection, are within the standard deviation of the surrounding speckle and thus unidentifiable in the resulting image. Our chosen example is therefore less tolerant to errors.

Another source of difficulty arises from our underlying assumption of the pulse shape. The assumed pulse model was developed for the main lobe reflection of the pulse, yet for more distant lines, representing side lobe echoes, the pulse shape is not guaranteed to remain undistorted.

Despite these limitations, our methods are successful at suppressing the side lobe artifacts.

\subsection{Cardiac Data Results} \label{section:cardiacResults}

In addition to synthetic phantoms, our methods were tested on several sets of consecutive frames of cardiac ultrasound data provided by GE Healthcare.

In this section we examine the results obtained by applying our methods to raw RF data acquired and stored for cardiac images of a healthy consenting volunteer. 
The acquisition was performed using a breadboard ultrasonic scanner employing a 64-element phased array probe. 
Operating in second harmonic imaging mode, pulses were transmitted at $1.7MHz$, and the
corresponding second harmonic signal, centered at $3.4MHz$, was then acquired. 
Data from all acquisition channels was sampled at $16MHz$ and collected along 120 beams, forming a $75^{\circ}$ sector.
The maximal imaging depth was $z=16cm$, implying a cycle time of $T=208\mu sec$. 
The imaging settings are summarized in Table~\ref{tab:cardiacParams}.

\begin {table}[ht]
\centering
\begin{tabular}{|c||c|} 
\hline 
  Parameter  & Value \\ 
\hline \hline
$c$ (Speed of Sound) & 1540 m/sec \\
\hline 
$f_0$ (Central Frequency) & 3.423 MHz \\ 
\hline 
$f_s$ (Sampling Frequency) & 16 MHz \\ 
\hline 
Element Width & 0.29 mm \\
\hline 
Number of Elements & 64 \\
\hline 
Number of Scan Lines & 120 \\
\hline 
Sector Size & 75 degrees \\ 
\hline
\end{tabular} 
\caption{Cardiac imaging parameters} \label{tab:cardiacParams}
\end{table}

First we want to evaluate the decomposition results. 
For this purpose, we compare the images reconstructed for the separated background components (without compression).

\begin{figure*}[!htb]
\centering
\subfloat[]{
\centering \includegraphics[scale=0.4,clip,trim=2cm 1cm 2cm 1cm]{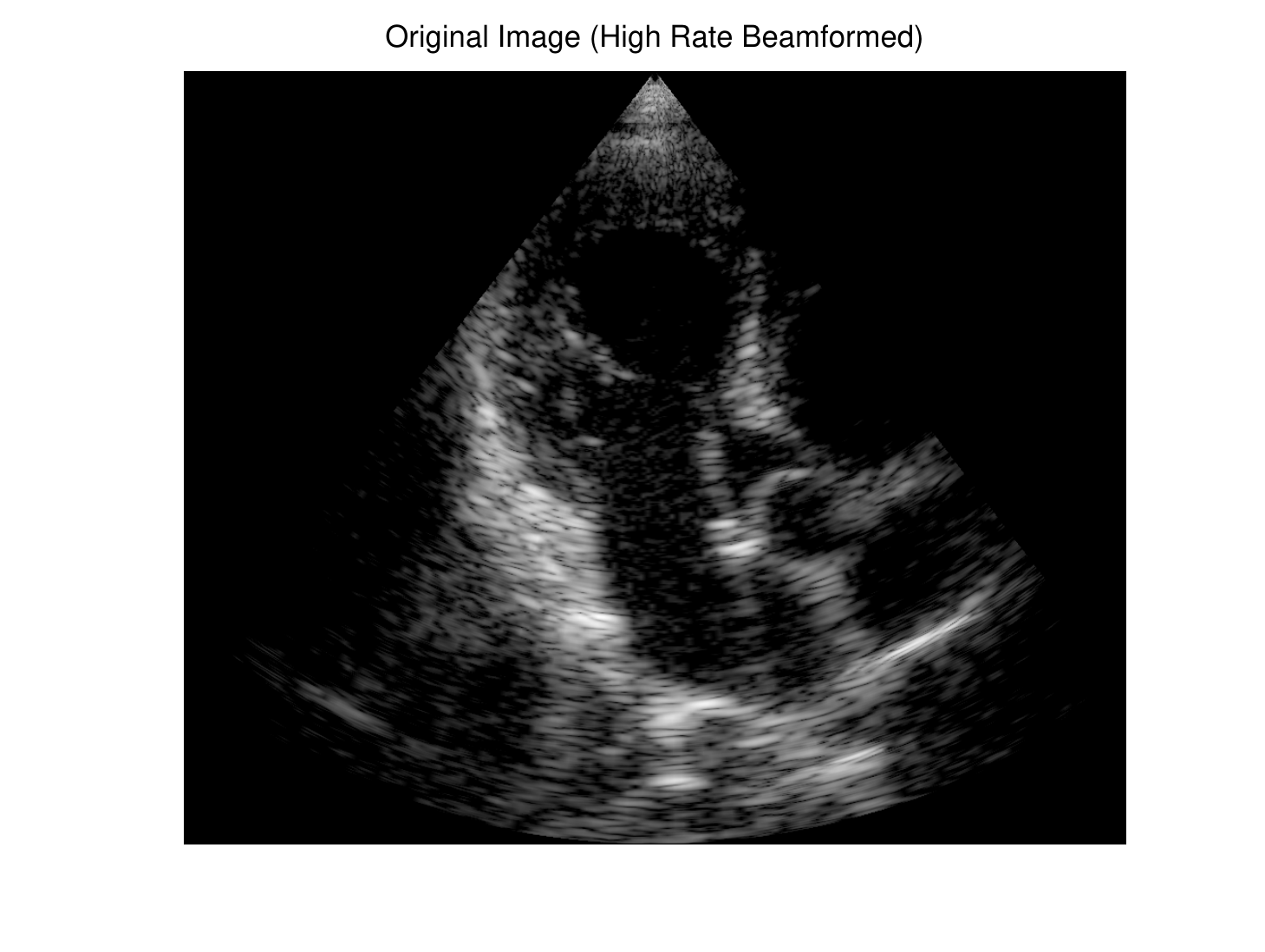}
\label{Fig:frame1a}
}
\subfloat[]{
\centering \includegraphics[scale=0.4,clip,trim=2cm 1cm 2cm 1cm]{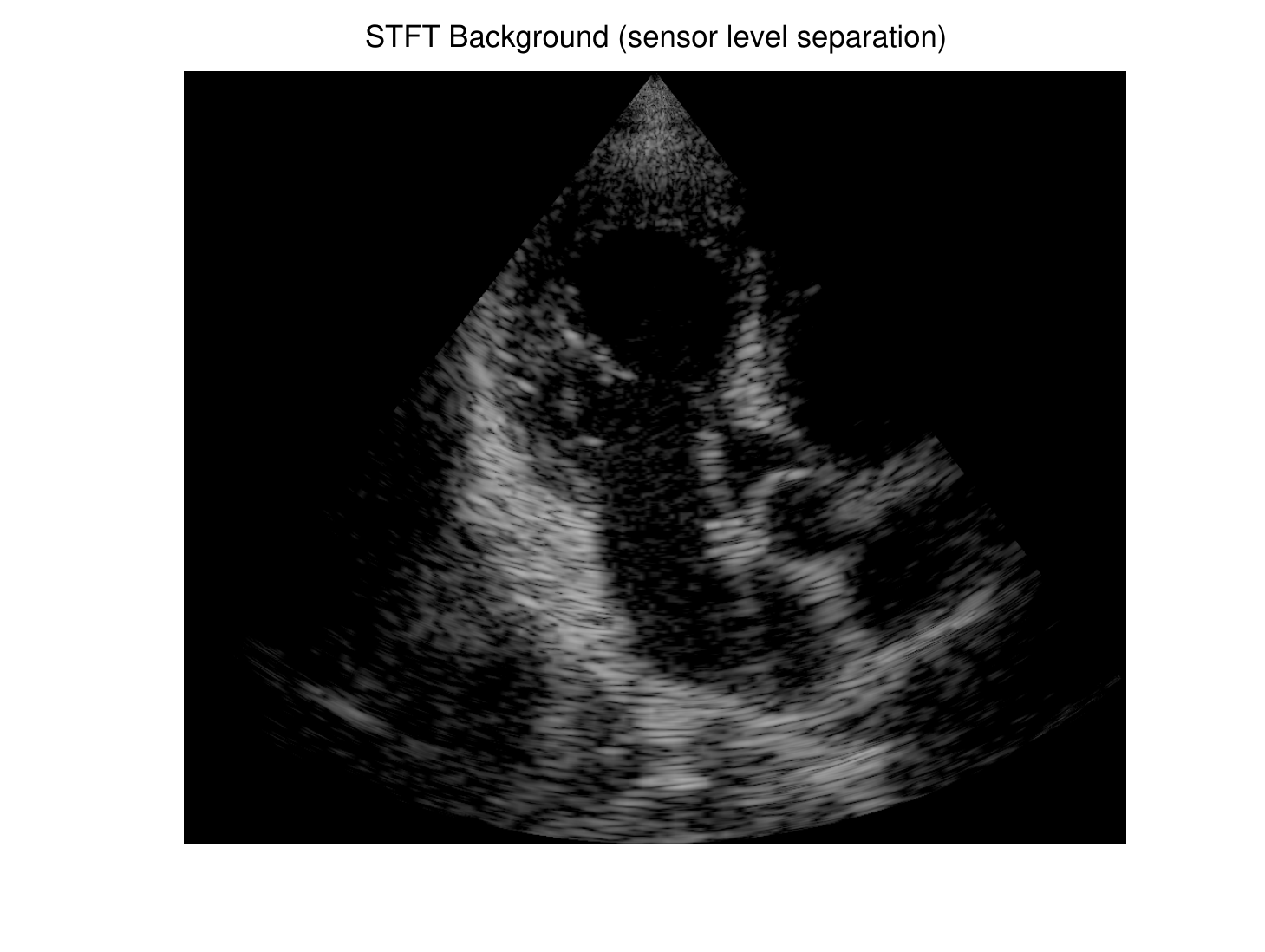}
\label{Fig:frame1b}
}
\subfloat[]{
\centering \includegraphics[scale=0.4,clip,trim=2cm 1cm 2cm 1cm]{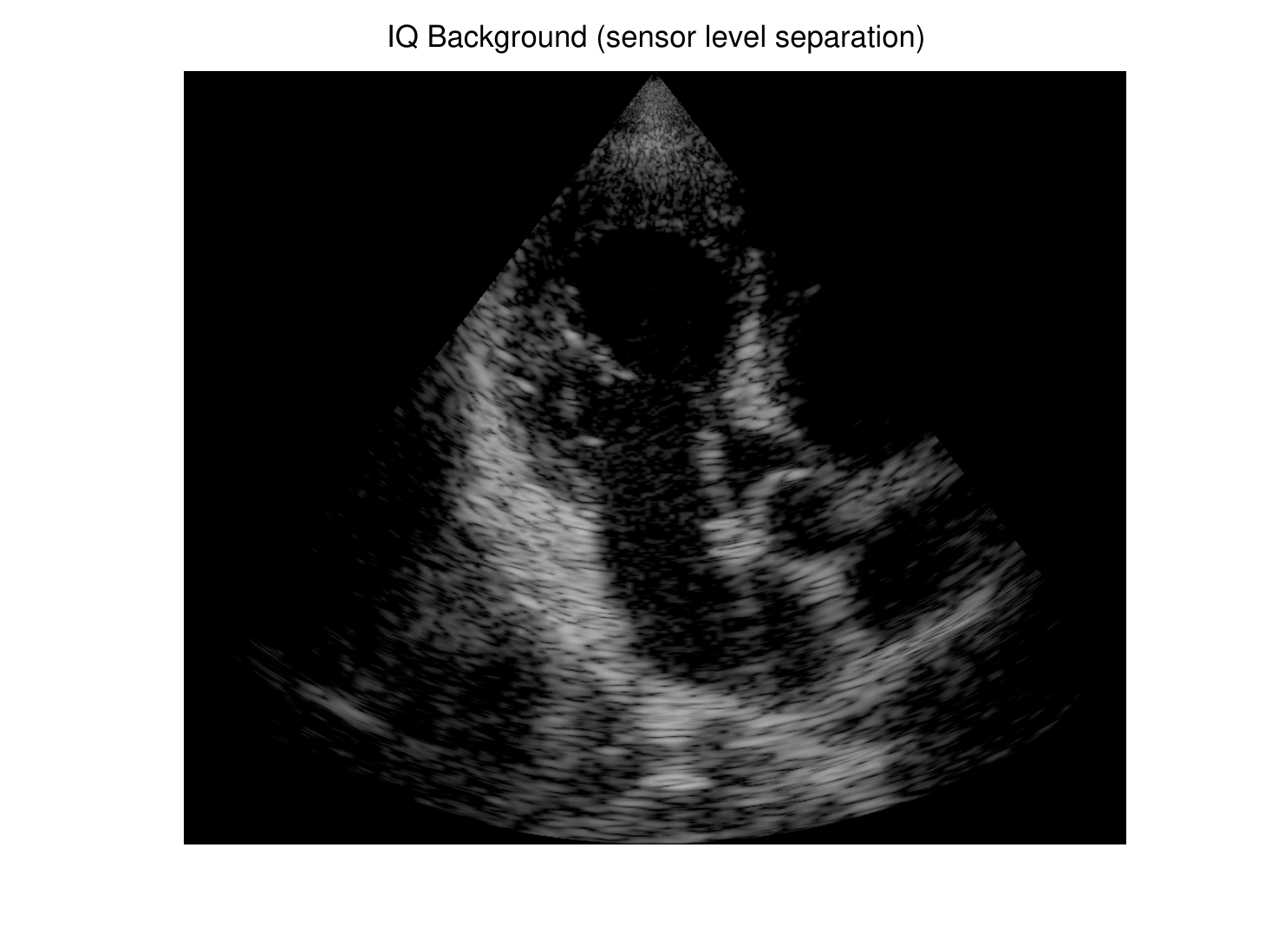}
\label{Fig:frame1c}
}\\
\subfloat[]{
\centering \includegraphics[scale=0.4,clip,trim=2cm 1cm 2cm 1cm]{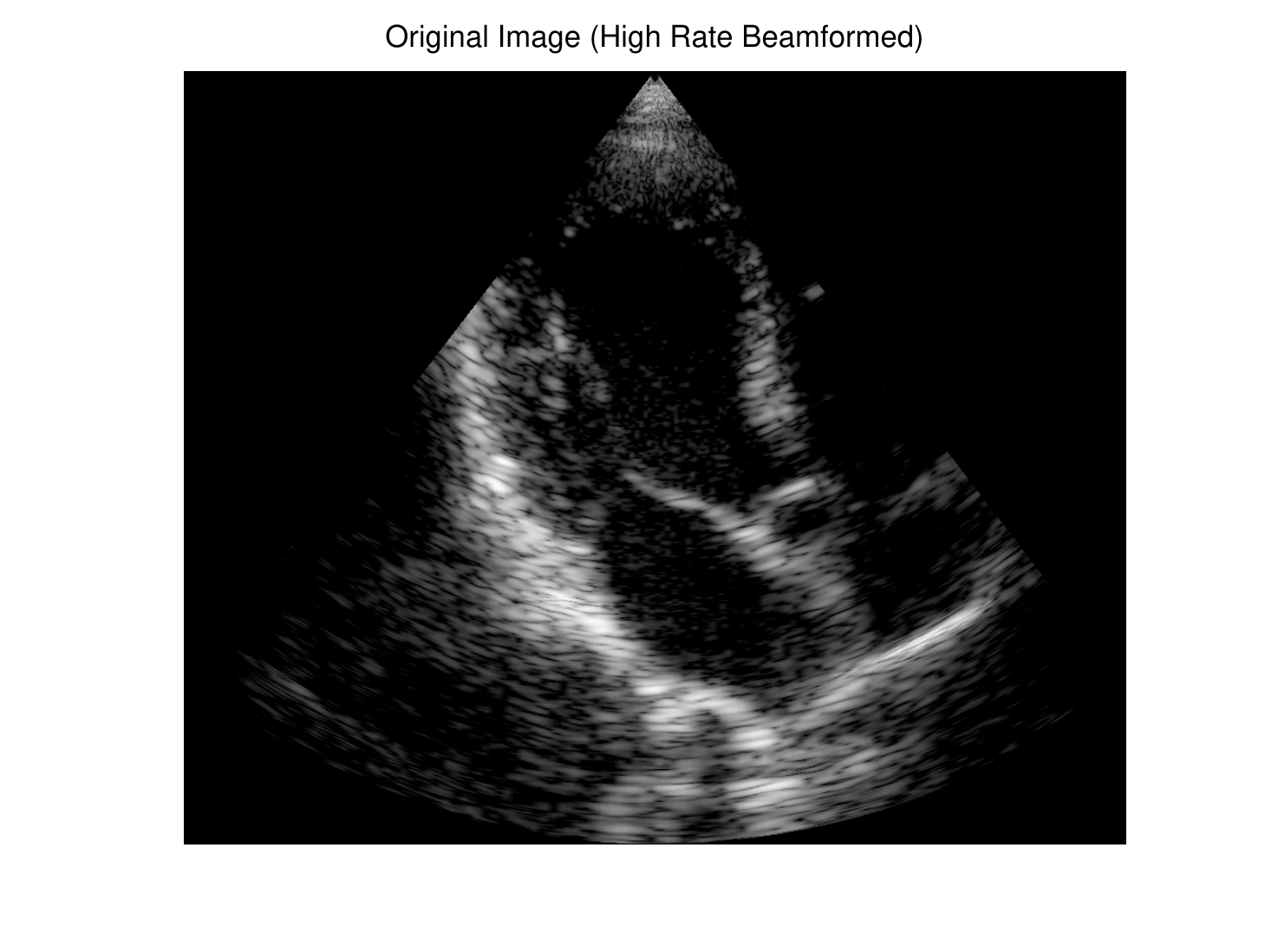}
\label{Fig:frame2a}
}
\subfloat[]{
\centering \includegraphics[scale=0.4,clip,trim=2cm 1cm 2cm 1cm]{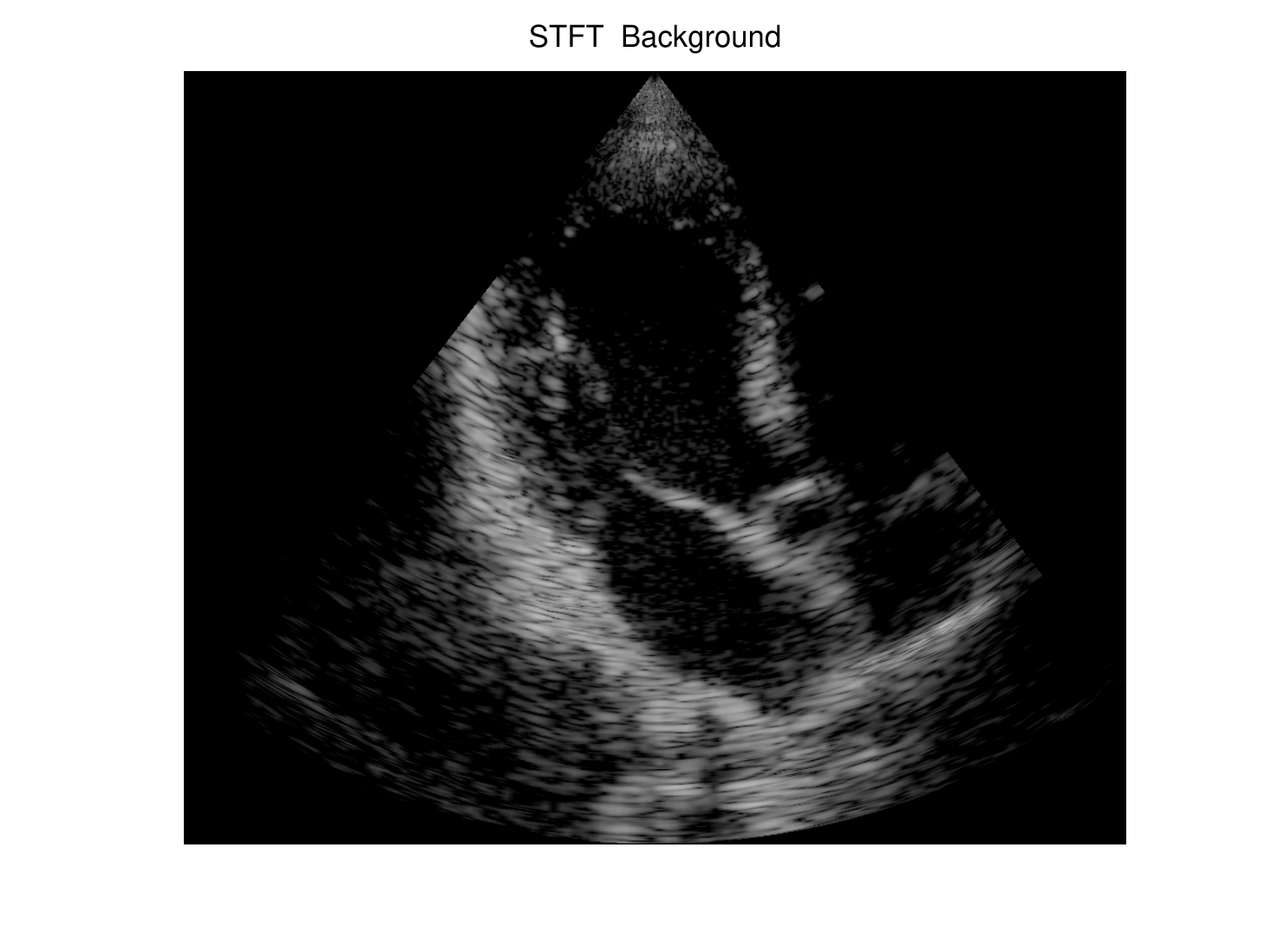}
\label{Fig:frame2b}
}
\subfloat[]{
\centering \includegraphics[scale=0.4,clip,trim=2cm 1cm 2cm 1cm]{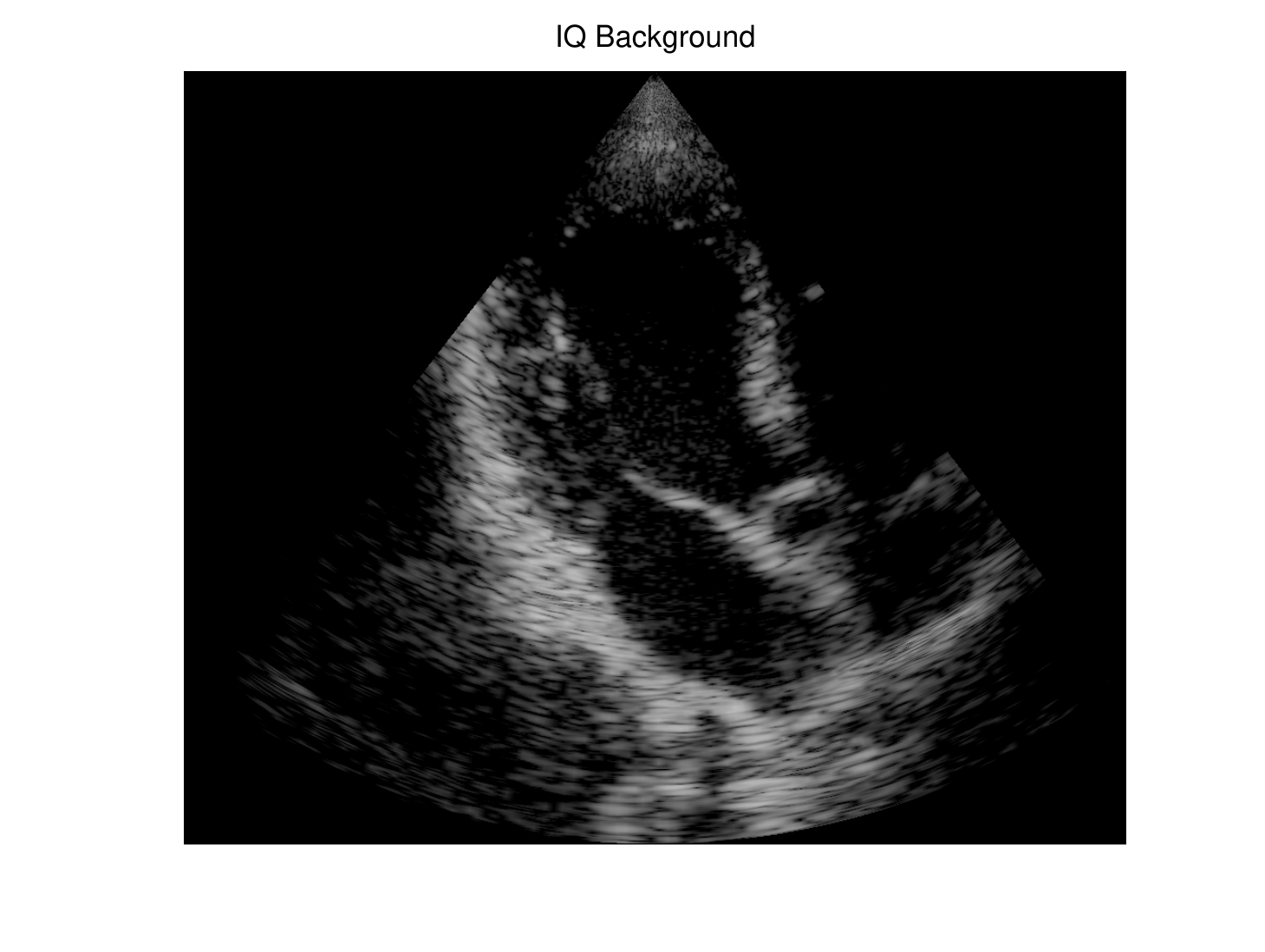}
\label{Fig:frame2c}
}
\caption[] {Background estimation results.
The first row corresponds to frame 1, and the second row corresponds to frame 2. \subref{Fig:frame1a},\subref{Fig:frame2a} Original image. \subref{Fig:frame1b},\subref{Fig:frame2b} STFT background image. \subref{Fig:frame1c},\subref{Fig:frame2c} IQ background image.}
\label{Fig:frames12}
\end{figure*}

The results obtained for 2 different 
frames are illustrated in Figure~\ref{Fig:frames12}, where each frame is presented along with its background estimations using the STFT and IQ based decomposition methods.
These images indicate that our proposed decomposition methods successfully detect and remove the strong reflections, thus producing a background image with relatively homogeneous regions. The two evaluated decomposition methods produce very similar results.
\newline

\begin{figure}[htb]
\centering
\subfloat[]{
	\centering
	\includegraphics[trim=2cm 1cm 2cm 1cm,clip,scale=0.4]{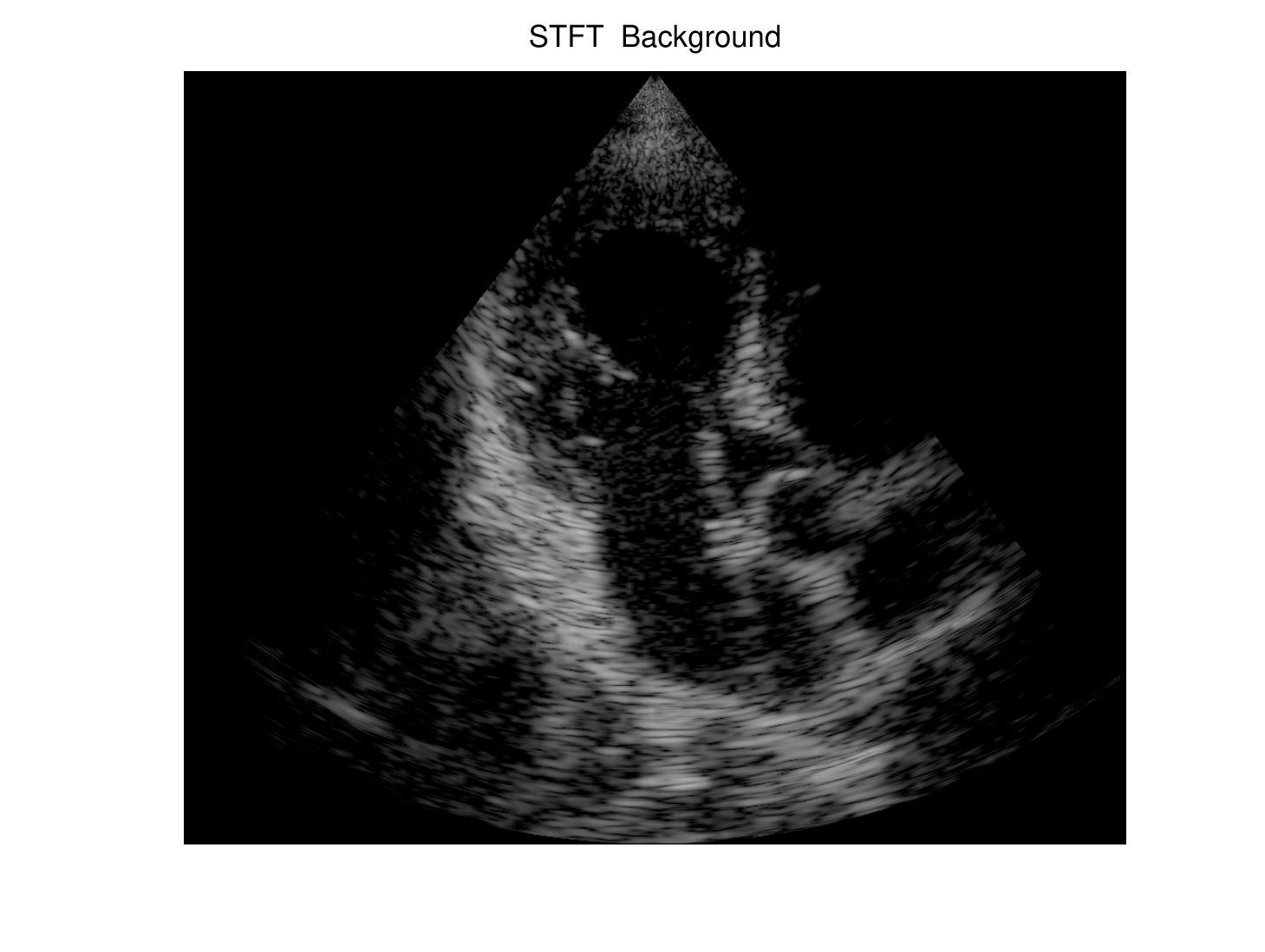}
	\label{Fig:STFT1_3e4}
}
\subfloat[]{
	\centering
	\includegraphics[trim=2cm 1cm 2cm 1cm,clip,scale=0.4]{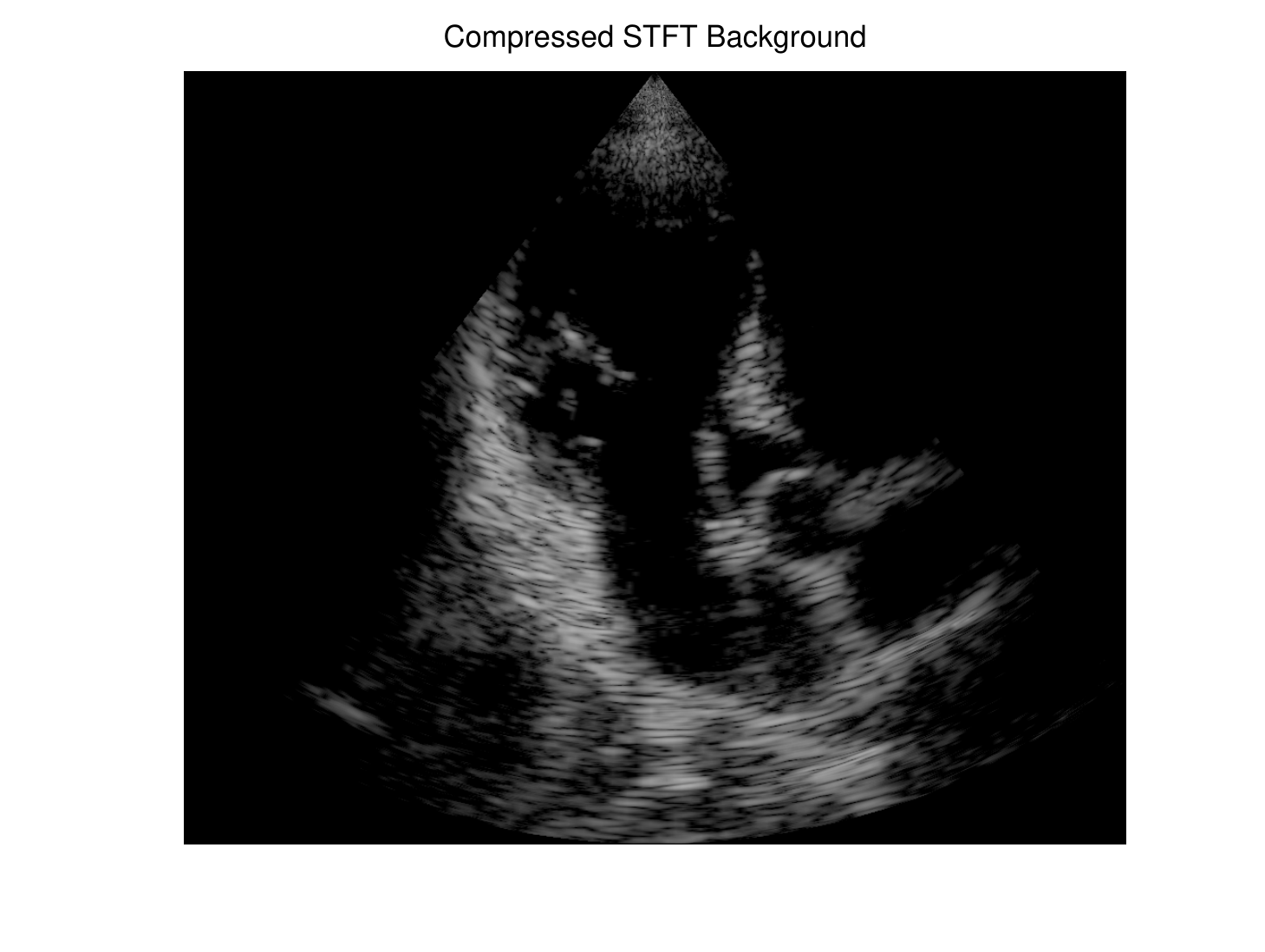} 
	\label{Fig:STFT1_sparse_3e4}
    \centering 
}
\caption[] {Frame 1 background compression results: \subref{Fig:STFT1_3e4} STFT background image, \subref{Fig:STFT1_sparse_3e4} Compressed STFT background image (PSNR=29.16[dB])}
\label{Fig:compressionResult_new1}
\end{figure}

\begin{figure}[!htb]
\centering
\subfloat[]{
	\centering
	\includegraphics[trim=2cm 1cm 2cm 1cm,clip,scale=0.4]{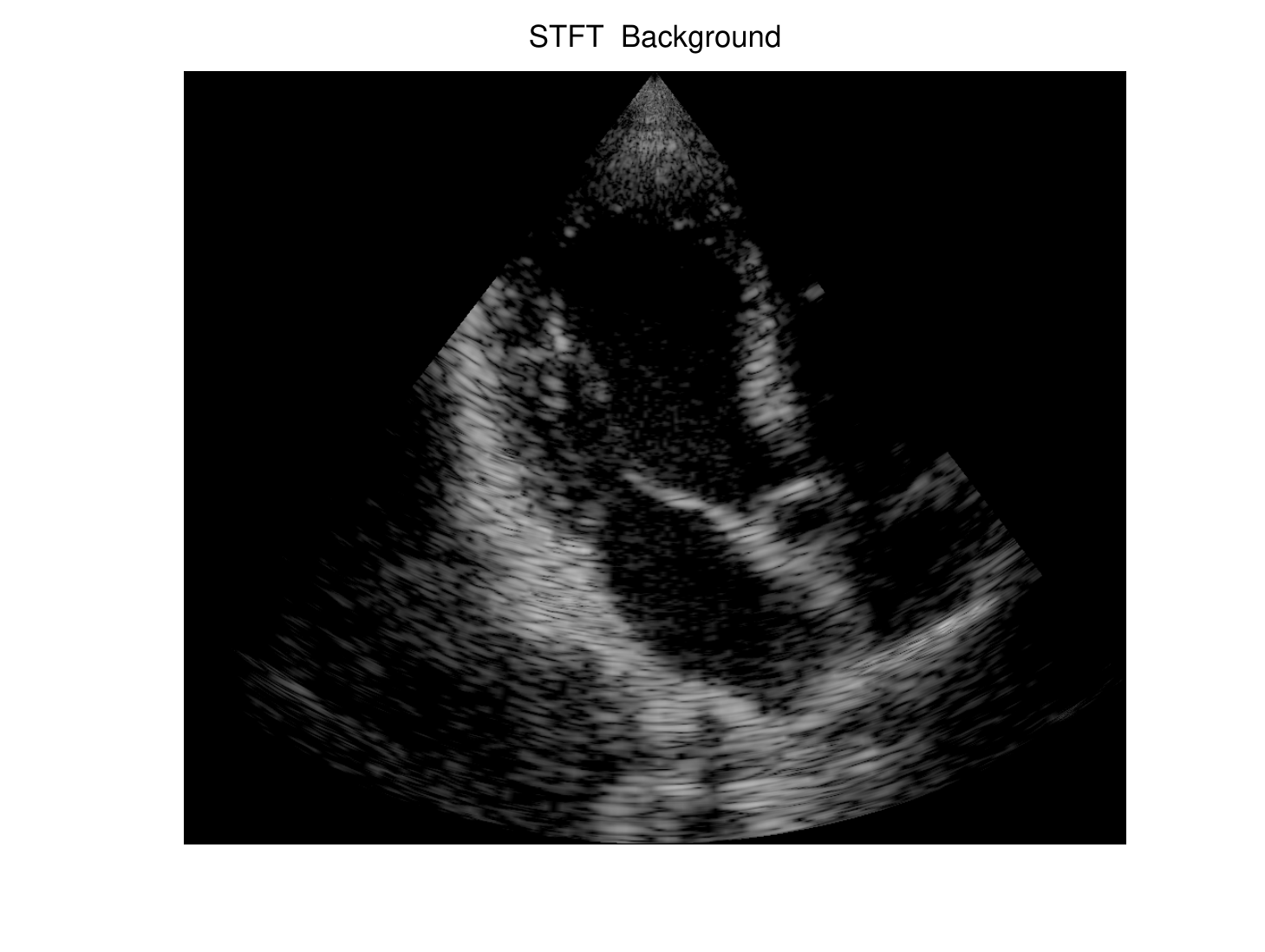}    	
	\label{Fig:STFT2_3e4}
}
\subfloat[]{
	\centering
	\includegraphics[trim=2cm 1cm 2cm 1cm,clip,scale=0.4]{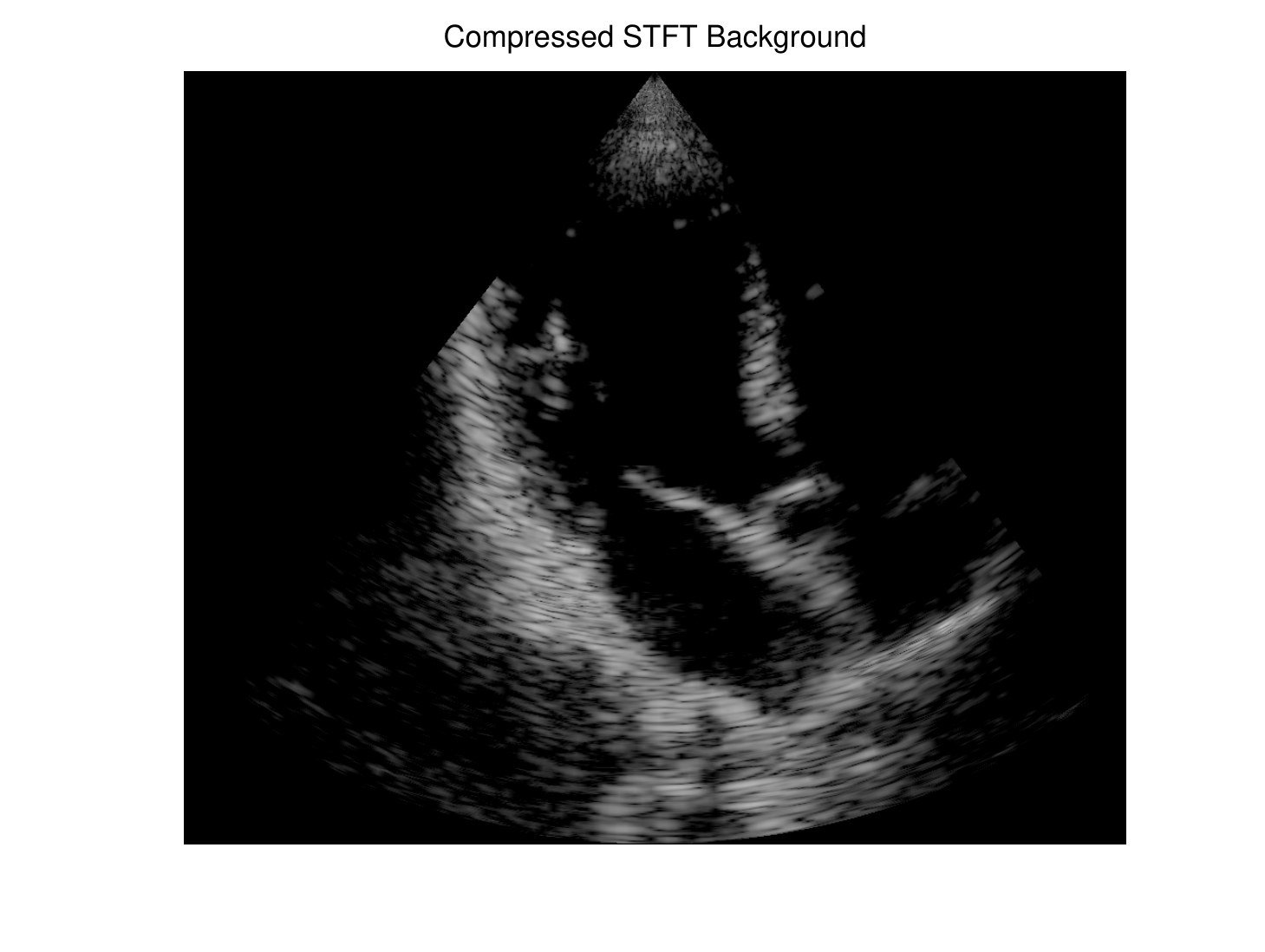} 
	\label{Fig:STFT2_sparse_3e4}
    \centering 
}
\caption[] {Frame 2 background compression results: \subref{Fig:STFT2_3e4} STFT background image, \subref{Fig:STFT2_sparse_3e4} Compressed STFT background image (PSNR=29.41[dB])}
\label{Fig:compressionResult_new2}
\end{figure}

Following decomposition, background compression was conducted according to the algorithm described in Section~\ref{section:BackgroundCompression}.
Non-overlapping patches of length $Q=100$ were used, the number of sensors is $M=64$ and the RF signal length is $N=3328$ samples.
It should also be noted that in this experiment, apodization was applied to each raw signal before summing them up, which is not reflected in the theoretical formulation of representation domain beamforming presented in Section~\ref{section:compressedBF}. 
The results are depicted in Figure~\ref{Fig:compressionResult_new1}, that presents the background image obtained using STFT based decomposition alongside its 24-fold compressed version, as applied to frame 1, whose data was used as the training set.
Figure~\ref{Fig:compressionResult_new2} then displays the compression results obtained for frame 2, which indicate that although the dictionary was trained based on signals from a single frame, it is suitable for representing data of other frames as well. 
Similar results were obtained for other sets of cardiac ultrasound frames acquired using the same imaging settings.

In Section~\ref{section:BeackgroundDictLearning} we pointed out a few challenges to the background compression process, namely the small size of the dictionary training set, the use of non-overlapping patches and the fact that training was performed using beamformed signals while intended to represent raw signals.

Observing the compressed images, our proposed compression scheme seems to produce visually good images that preserve even the subtle image features.
These results were obtained despite the formerly mentioned challenges, and while achieving an average compression ratio of 24, implying that the number of coefficients needed for reconstruction was only $4\%$ of the number of time samples in the received RF signal.
\newline

To further demonstrate the superiority of our chosen compression scheme, let us return to our initial conjecture motivating the raw signals decomposition, which was that individually compressing each component may lead to better compression ratios compared with a direct compression of the complete raw signal.
In order to prove this hypothesis, we conducted a dictionary learning process using the same methods previously applied for the background signals, now using the original beamformed signals as the training set, without removing the strong reflectors. 
We then used this new dictionary for sparsely representing each of the detected raw signals.

For the sake of comparison, the same error threshold was used for the dictionary learning process, and the learning set also consisted of 10 beamformed signals, taken from the same frame dataset.

\begin{figure}[ht]
\centering
\subfloat[]{
    \centering \includegraphics[trim=2cm 1cm 2cm 1cm,clip,scale=0.4]{frame1_orig}
	\label{Fig:frame1_orig}
}
\subfloat[]{
    \centering \includegraphics[trim=2cm 1cm 2cm 1cm,clip,scale=0.4]{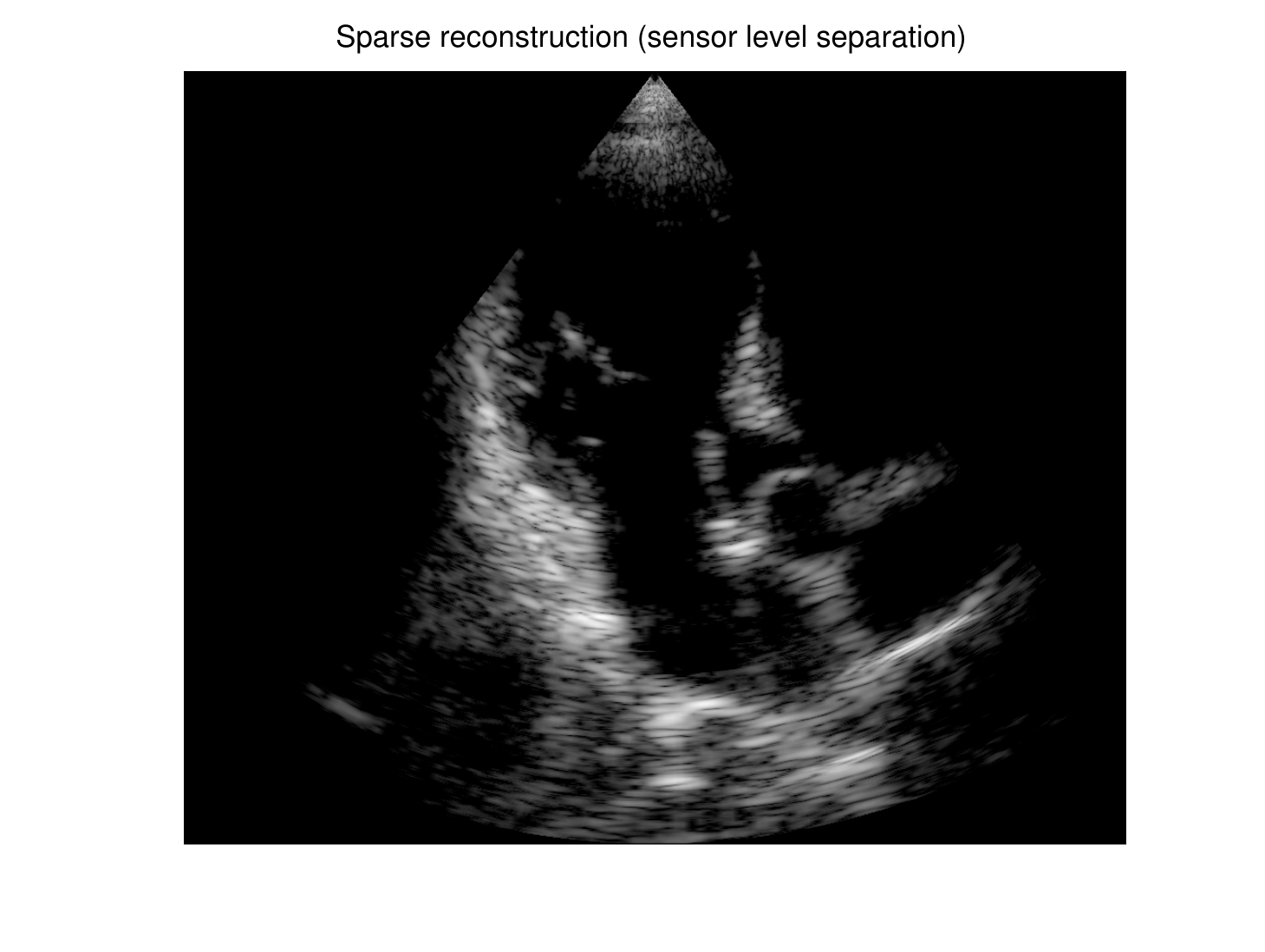}
	\label{Fig:orig1_sparse}
}
\caption[Compression results for frame 1 without decomposition] {Compression results for frame 1 without decomposition: \subref{Fig:frame1_orig} Original image, \subref{Fig:orig1_sparse} Compressed image.}
\label{Fig:OrigFrame1Compression}
\end{figure}

The compression results obtained for the original frame data (without decomposition) over this specifically trained dictionary, are presented in Figure~\ref{Fig:OrigFrame1Compression}.
The resulting image is of comparable quality to that of the compressed background images displayed in Figures~\ref{Fig:compressionResult_new1}-\ref{Fig:compressionResult_new2}, in terms of the visible amount of saved features.
The benefit though is in the achieved compressed ratio, which is over 2-fold better when utilizing decomposition.
\newline

Since the original image contains the strong reflectors information as well, a fair comparison of the compression ratios demands that the amount of coefficients needed for representing the strong reflectors is added to those used for representing the background signal.
When the strong reflectors are considered, a slightly reduced compression ratio of 21.4 is achieved. Nonetheless, this achieved compression ratio is still twice as high as the one achieved for the original raw data.
\newline

The comparison of achieved compression ratios is summarized in Table~\ref{tab:CompressionRatios}.
The first line presents information for the compression of the background only, the second presents information combined from separate compression of the background and the strong reflectors, and the third line presents information for the direct compression of the raw signals without decomposition.

\begin {table}[ht]
\centering
\begin{tabular}{|c||c|c|}  
\hline 
 & \% of coefficients &  compression factor  \\ 
\hline \hline
background  & 4.07 & 24.56 
\\ 
\hline 
\begin{tabular}{c} background \\ + strong reflectors \end{tabular} & 4.67 & 21.40 
\\
\hline
full raw signal & 9.10 & 10.99 
\\
\hline
\end{tabular}
\caption[Comparison of the achieved compression ratios]{Comparison of the achieved compression ratios} \label{tab:CompressionRatios}
\end{table}

It is readily seen that the individual compression of the background and strong reflectors components outperforms the direct compression of the raw data, in accordance with our predictions.
\newline

In this regard, it should be pointed out that the compression ratio considering both the background signal and strong reflectors, is relevant for analyzing the total amount of saved data. However, in terms of the data needed to be employed in beamforming computations, the higher compression factor (that only considers the background) is still applicable, as our proposed processing scheme suggests that the strong reflectors need not undergo beamforming for their reconstruction.
Moreover, recall that the beamforming process itself was simplified by conducting it in the representation domain, thus achieving further reduction of the computational load besides the use of a smaller amount of coefficients.
\newline

\section{Conclusions} \label{section:Conclusions}
In this work, we extended previous models proposed in \cite{Tur2010,Wagner2012} by integrating the speckle reflections and assembling a direct sum of two components, each of which carries valuable information and could be characterized by a limited amount of parameters.
In accordance with this model, we developed a processing scheme for raw ultrasound signals that exploits the inherent redundancy of the data, and achieves an improved compression ratio as well as a significant reduction of side lobes artifacts, while preserving the image information. 

At the heart of the proposed processing scheme stands a sparse decomposition stage, that detects the strong reflections and separates them from the background signal.
The first approach for doing so is based on the Short-Time Fourier Transform (STFT), such that decomposition is carried out in a time-frequency domain.
A second approach, adapted to state-of-the-art ultrasound systems in which IQ demodulation is performed prior to sampling, applies decomposition directly to the I/Q components.
We next provided slightly altered versions of both these approaches, that improve their ability to suppress reflections originating from side lobes.

After the signal decomposition stage, the separated background component undergoes further processing. First, it is sparsely represented over a suitable dictionary, that was trained offline from background signal examples. Afterwards, the compressed background signals are integrated in a modified beamforming process, applied at the representation domain.

Finally, the strong reflectors could be reconstructed from their sparse coefficients directly into the beamformed signal, and combined with the reconstructed background component in order to construct the complete signal.
\newline

The novelty of this model lies in the component-based approach, especially as it concerns the raw signals rather than the beamformed ones or the resulting image.

An important application that gains from this model and the derived processing scheme, is the reduction of the amount of data needed to be transferred from the system front-end and processed by the beamformer.
Applying our processing schemes to real cardiac ultrasound data, we successfully reconstruct the image contents while achieving over twenty-fold reduction of the data size. 
For comparison, this compression rate is twice as high as the one achieved by an equivalent compression of the raw signal without decomposition.

We point out that designed for the raw signals, the online algorithm may be applied at the sensors immediately after sampling, and does not require the full frame to be acquired.

Moreover, it is clearly desirable to compress the data as early in the processing chain as possible. As far as digital compression is concerned, our approach operates on raw signals "close to the source", i.e. immediately after sampling. Though not yet attempted in the scope of our work, we believe that utilizing the proposed two-component model and learned dictionary, a low rate sampling scheme can be established, such that our algorithm may be extended to the compressed sensing framework. 
Doing so, our results could be compared with other ultrasound compression techniques currently employed in the analog domain.

An additional contribution of our work relates to the resulting suppression of side lobes.
We show that by separating the strong reflectors at early stages of the imaging cycle, before the receive beamforming, side lobes artifacts are significantly reduced alongside the data size reduction, thus improving the contrast of the reconstructed image and its diagnostic value.

Finally, we note that the component-based modeling may open more possibilities for analyzing ultrasonic signals. While we identified two main components, other decomposition ideas may be investigated, such as separating the first- and second- harmonic echoes, or detecting more than two components related to various artifacts which require special processing.

\bibliographystyle{IEEEtran}
\bibliography{mybib}

\end{document}